\documentclass[11pt]{article}
\usepackage{graphicx}
\usepackage{psfrag}
\usepackage{epsfig} 
\usepackage{psfig} 
\usepackage{cite}
\begin{document}
\begin{flushright}
hep-th/0205084\\
\end{flushright}
\vskip 1cm

\begin{center}
{\Large {\bf On Brane World Cosmological Perturbations}}\\
\vskip 1.5cm
C\'edric Deffayet\footnote{cjd2@physics.nyu.edu}\\
{\it Center for Cosmology and Particle Physics,\\
Department of Physics, New York University,\\
4 Washington Place, New York, NY 10003, USA.}
\end{center}
\vskip 1cm
\noindent
\begin{center}
{\bf Abstract}
\end{center}
\vskip .3cm
We discuss the scalar cosmological perturbations 
in a 3-brane world
 with a 5D bulk.
We first show explicitly how the effective perturbed Einstein's equations on the brane (involving the {\it Weyl fluid}) are encoded into 
Mukohyama's master equation. We  give the relation between Mukohyama's master variable and the perturbations of the Weyl fluid,  we also 
discuss 
the relation between the former and  the perturbations of matter and induced metric on the brane. We show 
that one can obtain a boundary condition on the brane  for the master equation 
solely expressible in term of the master variable, in the case of 
 a perfect fluid with  
adiabatic perturbations  on a Randall-Sundrum (RS) or
Dvali-Gabadadze-Porrati (DGP) brane. This provides an easy way to
solve numerically for the evolution of the perturbations 
as well as should shed light on the various approximations done in the
literature to deal with the Weyl degrees of freedom.

\pagebreak

\newcommand{\beq}{\begin{eqnarray}}
\newcommand{\eeq}{\end{eqnarray}}
\newcommand{\kc}{\kappa_{(5)}}
\newcommand{\kq}{\kappa_{(4)}}
\newcommand{\kcd}{\kappa^2_{(5)}}
\newcommand{\kcq}{\kappa^4_{(5)}}
\newcommand{\kqd}{\kappa^2_{(4)}}
\newcommand{\Lc}{\Lambda_{(5)}}
\newcommand{\Lq}{\lambda_{(4)}}
\newcommand{\abd}{\dot{a}_{(b)}}
\newcommand{\abp}{a^\prime_{(b)}}
\newcommand{\ab}{a_{(b)}}
\newcommand{\nbd}{\dot{n}_{(b)}}
\newcommand{\nbp}{n^\prime_{(b)}}
\newcommand{\nb}{n_{(b)}}
\newcommand{\bb}{{(b)}}
\newcommand{\MM}{{(M)}}
\newcommand{\EE}{{({\cal E})}}
\newcommand{\hd}{\dot{H}}
\newcommand{\hdd}{\ddot{H}}
\newcommand{\hddd} {H^{\cdot \cdot \cdot}}
\newcommand{\da}{\dot{a}}
\newcommand{\db}{\dot{b}}
\newcommand{\dn}{\dot{n}}
\newcommand{\dda}{\ddot{a}}
\newcommand{\ddb}{\ddot{b}}
\newcommand{\ddn}{\ddot{n}}
\newcommand{\paBDL}{a^{\prime}}
\newcommand{\pb}{b^{\prime}}
\newcommand{\pn}{n^{\prime}}
\newcommand{\ppa}{a^{\prime \prime}}
\newcommand{\ppb}{b^{\prime \prime}}
\newcommand{\ppn}{n^{\prime \prime}}
\newcommand{\fda}{\frac{\da}{a}}
\newcommand{\fdb}{\frac{\db}{b}}
\newcommand{\fdn}{\frac{\dn}{n}}
\newcommand{\fdda}{\frac{\dda}{a}}
\newcommand{\fddb}{\frac{\ddb}{b}}
\newcommand{\fddn}{\frac{\ddn}{n}}
\newcommand{\fpa}{\frac{\paBDL}{a}}
\newcommand{\fpb}{\frac{\pb}{b}}
\newcommand{\fpn}{\frac{\pn}{n}}
\newcommand{\fppa}{\frac{\ppa}{a}}
\newcommand{\fppb}{\frac{\ppb}{b}}
\newcommand{\fppn}{\frac{\ppn}{n}}
\newcommand{\UU}{\Upsilon}
\newcommand{\fU}{f_\UU}
\newcommand{\CC}{{\cal C}}

\section{Introduction}
The phenomenology of brane world models has been the subject of intensive investigations in the last years. 
Its richness is in part due to the fact that those models can lead to modifications of gravity at small, but almost macroscopic,  or even very large (cosmological) distances. 
The first case occurs  in  e.g.  Arkani-Hamed-Dimopoulos-Dvali  \cite{Arkani-Hamed:1998rs}  or Randall-Sundrum  \cite{Randall:1999vf}  type of models,  
where the recovery of ordinary gravity at every-day scales is due to the geometry of the bulk space time, and gravity gets modified when Kaluza-Klein modes become excited.  On the other hand, the Dvali-Gabadadze-Porrati (referred to as the DGP model in the following) brane worlds model \cite{DGP} leads to  modification of gravity at very large (cosmological or larger) distances, in addition to possible small distance modification of different nature (see \cite{DGP,DG,Dvali:2001gm,Dvali:2001gx}) .  
In this  model, the  recovery of 4D gravity for usual distances 
 does not rely on the geometry of the bulk, but 
is ensured by 
the effect of  an induced Einstein-Hilbert term in the action, while the extra dimension can be non compact and have {\it infinite volume}
\footnote{This very fact can lead to new ways of addressing the cosmological constant problem \cite{DGP,Witten:2000zk}}. 
Because of those modifications of gravity, the cosmology of brane worlds can differ dramatically from standard cosmology  
and can potentially lead to various ways to test them. This is particularly true with the advent of precision cosmological measurement. Conversely the brane-world models can also lead to  new scenarios for the primordial universe (like e.g. in  \cite{tot}) or  its recent cosmological evolution (as in 
the DGP model, which  has  the ability to produce acceleration of the universe, as suggested by SNIa data \cite{cc},
 without the need for a non zero cosmological constant \cite{Deffayet:2001uy,Fifth} in a way currently in agreement with   
supernovae and Cosmic Microwave Background (CMB) data \cite{Deffayet:2002sp}\footnote{As far as the CMB is concerned, the model has only been tested in \cite{Deffayet:2002sp}  assuming standard growth of cosmological perturbations, but non standard background evolution. One of the purpose of this work is to develop the formalism required to go beyond 
 this approximation
and test its validity. We will however, for the sake of simplicity,
only give here the equations for a cosmological DGP brane
which bulk space-time lies in its interior, which does not correspond to the case considered in 
\cite{Fifth} (see \cite{Deffayet:2001uy}).} (see also \cite{Avelino,comment})). The DGP model is also 
 in itself an interesting playground to investigate  the van Dam-Veltman-Zakharov discontinuity \cite{Veltman}.

Measurement of the anisotropies of the Cosmic Microwave Background, and large scale galaxies or weak lensing surveys 
provide a unique way to test gravity at large scales (see e.g. \cite{LS}) but also our 
comprehension of the physics of the primordial universe. 
As far as brane world models are concerned, this  relies in particular on a better understanding of cosmological perturbations in these kind of models. 
The present article goes along this line and aim at reporting progresses in the study of brane  world cosmological perturbations. The latter 
 have already been investigated  by many authors
\cite{GS,Deruelle,Langlois,Bridgman:2001mc,Langlois:2001ph,Roy1,LMSW,Mukohyama:2000ui,Mukoh3,Leong:2001qm,cvdb,Mukoh2} with only limited results, as far as observable predictions are concerned (see e.g. \cite{Bridgman:2001mc,review}). 
We will only consider here the case of scalar perturbations 
in a brane world model with a single 3-brane of codimension one.
The bulk space-time will not be assumed to be {\it stabilized}, which is well suited for the DGP model or the Randall-Sundrum {\it II} model (presented in \cite{Randall:1999vf}, that we will call the RS model in the following).  
In the latter case, it has been shown possible to obtain effective local perturbed Einstein's equations on the brane, which can be easily compared with usual  perturbed 4D Einstein's equations \cite{Langlois:2001ph}.
Apart from the different time evolution of the background, 
and terms arising from the brane typical quadratic coupling between brane matter and induced metric \cite{Binetruy:2000ut},
the extra-dimension manifests itself by source terms in the effective Einstein's equations which are coming from the bulk Weyl's tensor
\cite{Shiromizu:2000wj} and represent the imprint of bulk
gravitational waves on the brane. Those equations do not close on the
brane  because of the lack of a local evolution equation for all the
Weyl degrees of freedom\footnote{namely, for the Weyl's fluid anisotropic stress.}. It is however possible to close the system for large scale cosmological perturbations \cite{Roy1}, but this does not enable to compute the Sachs-Wolfe effect \cite{LMSW}. 
In order to obtain the evolution equations for cosmological perturbations on the brane, one should then solve the equations of motions for perturbations in the bulk. In this perspective, one may wish to use the work of Mukoyama who showed that the bulk scalar Einstein's equations are solved by a single master variable obeying a master equation \cite{Mukohyama:2000ui}.
The master variable  can be related to brane matter by junction conditions on the brane \cite{Mukoh3}. One of the purpose of the work
reported here is to discuss the relationship between this approach and the previous one, and in addition the nature of the boundary condition for the master variable on the brane. This can ease the formulation of the problem if one wants to integrate numerically the equations of motion in the bulk, but also shed light on various approximations done in the literature to deal with the Weyl's degrees of freedom
\cite{LMSW,Leong:2001qm} as well as on the well-poseness problem. Although most of this work can be applied to any kind of model involving a 3-brane of codimension one embedded in a bulk with a cosmological constant, we will, when considering realization of this setup, only deal with the case of  RS and DGP model.

The paper is organized as follows.  In the rest of this section, we briefly introduce the models we are interested in and their geometry (subsections \ref{MOGEO} and \ref{GEOM}), and then review some properties of the background cosmological solutions for those models (subsection \ref{COSBACK}). We then introduce our notations for metric,  matter and Weyl's perturbations (respectively section \ref{METPER} and \ref{MATPER}), we also discuss causal properties of the master equation in relation with the issue of initial and boundary conditions (\ref{causmast}), both for RS and DGP model. We then show explicitly, in section \ref{PertEin},  how all the perturbed effective Einstein's equations on the brane, and conservation equations, are encoded into  the master equation. In section \ref{BOUND},  we obtain a boundary condition on the brane for the master variable for a perfect fluid on a RS or DGP brane, we also give the relation between the Weyl's fluid perturbations and the master variable.

\subsection{Models for brane worlds} \label{MOGEO}
We consider a 3-brane embedded in a 5D space-time. 
The bulk space-time will be considered  empty with a, possibly non vanishing, cosmological constant $\Lc$. \label{PAP-PCGI-1 p1+} 
The brane is defined as a hyper-surface $X^A(x^\mu)$, where $X^A$ are bulk coordinates, and 
$x^\mu$ are coordinates along the brane world-volume\footnote{Throughout this article, we will adopt the following  convention for indices:
upper case Latin letters $A,B,...$ will denote 5D indices: $0,1,2,3,5$; 
Greek letters  $\mu,\nu,...$ 
 will denote 4D indices parallel to the brane: $0,1,2,3$;
lower case Latin letters from the middle of the
alphabet: $i,j,...,$ will denote space-like 3D indices parallel to the brane: $1,2,3$.}. 
The bulk metric\  $g^{(5)}_{AB}$ induces through the brane embedding a metric on the brane,
 called the induced metric\footnote{We will in the following use 
an index $^{(4)}$ to distinguish, when necessary, a tensor depending only on the induced metric (such as the induced metric itself $g^{(4)}_{\mu \nu}$, or the Ricci scalar computed with the induced metric and covariant derivatives compatible with the latter, $R_{(4)}$) from its bulk counterpart (such as the bulk metric or the bulk Ricci scalar) to which we put an index $^{(5)}$.} and given  by 
\beq
\label{induite}
g^{(4)}_{\mu \nu} = \partial_\mu X^A \partial_\nu X^B g^{(5)}_{AB}.
\eeq
The gravitational part of the action of models under consideration can be taken as containing 
the 5D Einstein-Hilbert action given by 
\beq \label{S5}
S_{(5)}=-\frac{1}{2 \kcd} \int d^5X \sqrt{g_{(5)}} \left(R_{(5)} -2 \Lc \right),
\eeq
where $\Lc$ is the bulk cosmological constant,  $R_{(5)}$ is the 5D Ricci scalar, 
and $\kcd$ is the inverse  third power of the reduced 5D Planck mass. The models which 
we are interested in are (and have to be, in order to give an acceptable phenomenology) 
 all sufficiently close to ordinary 4D GR in some region of their phase space. For that reason we also introduce a dimension-full quantity $\kqd$ which we set to be the inverse square of the 4D reduced Planck mass.  

To account for the brane, we add to the action (\ref{S5}) a term taking care of   
 brane-localized fields given by 
\beq \label{S4}
S_{(4)}=\int d^4 x \sqrt{g_{(4)}} {\cal L},
\eeq
where ${\cal L}$ is a Lagrangian density that will be discussed below (we also implicitly include a suitable Gibbons Hawking term). 

It is always possible to choose a, so called Gaussian Normal (referred to as GN in the rest of this work), coordinate system where the line element 
can be put in the form 
\beq \label{GN}
ds^2 = dy^2 + g^{(4)}_{\mu \nu} dx^\mu dx^\nu, 
\eeq
and the brane is the hyper-surface defined by $y=0$.
In this coordinate system the gravitational  equations of motion derived from the sum of action 
(\ref{S5}) and (\ref{S4}) are simply given by 
\beq \label{einstein}
G^{(5)}_{AB} = \kcd \delta (y) S_{\mu \nu} \delta_A^\mu \delta_B^\nu - \Lc g^{(5)}_{AB}, 
\eeq
where $G^{(5)}_{AB}$ is the 5D Einstein's tensor, and 
$S_{\mu \nu}$ is the effective energy momentum tensor for the brane.
The latter is defined by the usual expression (where removing of total derivative terms is implicit) 
\beq \nonumber
S_{\mu \nu} = \frac{2}{\sqrt{g_{(4)}}} \frac{\delta}{\delta g_{(4)}^{\mu \nu}} \left(\sqrt{g_{(4)}} {\cal L} \right).
\eeq
In the following we will refer to this tensor as the {\it effective matter} energy momentum tensor to distinguish it from {\it real} matter energy momentum tensor (see below). For most the calculations done in sections
\ref{METPER}, \ref{MATPER} and \ref{PertEin}, we will not need to specify the explicit form of ${\cal L}$, which is model dependent. 
When dealing with existing models, we will consider two cases: 

(I) The Randall-Sundrum model  {\it II} (RS model in the following) \cite{Randall:1999vf},
 where the bulk is a slice of AdS$_5$ space-time\footnote{We will not discuss here the case of a brane in AdS-Schwarzschild, but a similar work is expected to apply to this case as well.}
of radius $r_{(5)}$ given by 
\beq \nonumber
r_{(5)} = \sqrt{-6 \Lambda_{(5)}^{-1}}.
\eeq
and  the brane is endowed with a tension (or cosmological constant)  $\Lq$, and brane 
localized matter (that we will call {\it real} matter in the following) of Lagrangian density ${\cal L}_{(M)}$.  In this case one has 
\beq \nonumber
{\cal L} = {\cal L}_{(M)} + \Lq.
\eeq
The brane tension and bulk cosmological constant fulfill the constraint 
\beq \nonumber
\Lc= - \kcq \frac{\Lq^2}{6}.
\eeq
The effective brane energy momentum tensor is given by 
\beq \label{SRS}
S_{\mu \nu} = T^{(M)}_{\mu \nu} -  \Lq g^{(4)}_{\mu \nu},
\eeq
where $T^{(M)}_{\mu \nu}$ is the {\it real} matter energy momentum
tensor. The Kaluza-Klein (KK) spectrum of gravitons in the RS model
contains a normalizable zero mode localized on the brane which
dominates at large distance the gravitational exchange between
sources. This gives rise to ordinary gravity at large distance, with a coupling given by 
\beq \nonumber
\kqd = \frac{\kcq \Lq}{6}.
\eeq
Below distances of order $r_{(5)}$, on the other hand gravity gets modified by massive KK 
excitations. 
 
(II) The brane-induced gravity model of Dvali-Gabadadze-Porrati
 (DGP model in the following) \cite{DGP}, where the bulk will be taken to be 
 a slice of Minkowski$_5$ space-time\footnote{See \cite{Dvali:2001gm} for discussions of more complicated bulk geometry.} (with vanishing cosmological constant and a vanishing Weyl's tensor)  and 
an Einstein-Hilbert term computed with the induced metric on the brane is present in the brane action. 
The brane Lagrangian density is given by 
\beq \nonumber
{\cal L} = {\cal L}_{(M)} - \frac{1}{2 \kqd} R^{(4)}.
\eeq
In this case $S_{\mu \nu}$ is given by 
\beq \label{SDGP}
S_{\mu \nu} = T^{(M)}_{\mu \nu} -  \frac{1}{\kqd} G^{(4)}_{\mu \nu}. 
\eeq
\label{PCGI25 p13}
In the DGP model,  the gravitational potential between static point-like sources 
behaves as in 4D (varying as the inverse distance between sources) for inter-source distances much 
 smaller than a crossover distance $r_c$ given by 
\beq
r_c = \frac{\kcd}{2 \kqd}, \label{rc}
\eeq
while it behaves as a 5D potential (varying as the inverse distance squared) 
for much larger distances.  
As mentioned in the previous subsection, one also expects to get small 
scales modifications of gravity, coming from quantum gravity effects, at distances smaller than $M_{(5)}^{-1}$ \cite{Dvali:2001gx}. This
goes beyond the classical analysis of this paper and will not be discussed in this work.
In the DGP  model, there is no normalizable zero mode for the graviton excitation, and gravity, from a 4D point of view, is transported by massive KK modes.
This can be understood as the source of the van Dam-Veltman-Zakharov discontinuity
appearing in the DGP model in the linearized theory over a flat
(Minkowski$_4$) brane \cite{DGP}. We will not discuss this issue here
and refer the reader to Refs. \cite{DGP,Deffayet:2002uk,Lue:2001gc, Gruzinov:2001hp,Porrati:2002cp}.

\subsection{Some geometry of brane world} \label{GEOM} 
Einstein's equations (\ref{einstein})  lead to Israel's Junctions conditions which 
relate the jump of the extrinsic curvature tensor of the brane $K_{\mu \nu}$ to whatever distributional source 
appears on the r.h.s. of  (\ref{einstein}) (here accounting for the
brane localized fields or induced metric dependent terms). 
When a Gaussian normal coordinate system (\ref{GN}) is chosen, 
the extrinsic curvature tensor of a given $y=constant$ hyper-surface is defined by 
\beq \label{junc1}
K_{\mu \nu} = \frac{1}{2} \partial_y g^{(4)}_{\mu \nu}. 
\eeq
Israel's Junction conditions 
derived from (\ref{einstein})
simply read in this case 
\beq \label{Israel}
K_{\mu \nu}^\bb = -\kcd \left(\frac{1}{2}  S_{\mu \nu} - \frac{1}{6} S g^{(4)}_{\mu \nu} \right),
\eeq
where $S \equiv  S_{\mu \nu} g_{(4)}^{\mu \nu}$ is the trace of the effective 
energy momentum tensor of the brane, 
 we have assumed  a brane with  $Z_2$ symmetry,
and the index $_\bb$ means accordingly\footnote{When dealing with tensors, space-time indices will always be written without parenthesis, while other indices referring to properties of the quantity considered, such as $^\MM$, $^\bb$, $^{(4)}$, $^{(5)}$ will always be parenthesized.} 
that the value of the component (here, as for any expression in the rest of this paper)  $K_{\mu \nu}$
is taken on one side of the brane, namely in $y=0^+$. 
The above equation  (\ref{Israel}) can also be rewritten 
\beq \nonumber
\left\{K_{\mu \nu} - K g^{(4)}_{\mu \nu}\right\}_\bb = - \frac{1}{2} \kcd S_{\mu \nu},
\eeq
where $K$ is defined by $K \equiv K_{\mu \nu} g_{(4)}^{\mu \nu}$. 

Using a Gauss decomposition
and the above Israel's conditions, 
one can then derive effective 4D Einstein's equations
on the brane relating the brane intrinsic curvature to its effective energy momentum content  \cite{Binetruy:2000ut}.
In particular one gets the following equation \cite{Shiromizu:2000wj}
\beq \label{effein}
G^{(4)}_{\mu \nu} = -\frac{1}{2} g^{(4)}_{\mu \nu}  \Lc +  \kcq \Pi_{\mu \nu} - {\cal E}_{\mu \nu},
\eeq
where $G^{(4)}_{\mu \nu}$ is the 4D Einstein tensor,  $ \Pi_{\mu \nu}$ is given by 
\begin{eqnarray} \nonumber
\Pi_{\mu \nu} &=& -\frac{1}{4} {S}_{\alpha \mu}  {S}^{\alpha}_\nu + \frac{1}{12} {S} {S}_{\mu \nu} + \frac{1}{8} 
{S}_{\alpha \beta} 
{S}^{\alpha \beta} g^{(4)}_{\mu \nu} - \frac{1}{24} {S}^2 g^{(4)}_{\mu \nu},
\end{eqnarray}
and ${\cal E}_{\mu \nu}$ is defined as the limiting value on the brane of the electric part of the bulk Weyl's tensor. In the Gaussian normal coordinate system (\ref{GN}) it is simply given by \label{PCGI-9 p1}
\beq \label{Weyldef}
{\cal E}_{\mu \nu} = C^{5}_{\quad \mu 5 \nu \bb},
\eeq
where $C^{A}_{\quad BCD}$ is the bulk Weyl's tensor. One sees that $ {\cal E}_{\mu \nu}$
acts in the effective Einstein's equations 
 (\ref{effein}) as an external source with an energy momentum tensor $T^\EE_{\mu \nu}$ that one can define as 
\beq \label{defTW}
T^\EE_{\mu \nu} = - \frac{1}{\kqd} {\cal E}_{\mu \nu}.
\eeq Following previous works, we will refer   to this source as the {\it Weyl's fluid} on the brane. 
Because the trace of the Weyl's tensor vanishes, one sees easily 
from the definition (\ref{defTW}) 
\label{PCGI 23 p 29-30} that 
in GN coordinate system (\ref{GN}) one  has 
\beq \label{traceW}
T^\EE \equiv T^{\EE \mu}_\mu = 0, 
\eeq
so that the Weyl's fluid shares some similarities with a radiation fluid.
An other useful identity is the Codacci equation reading  in a GN coordinate system 
\beq \nonumber
D_\nu K^{\nu}_{\mu} - D_\mu K = R^{(5)}_{y  \mu},
\eeq 
where $D_\mu$ is the covariant derivative compatible with the induced metric. 
Using the limiting value of the Einstein's equations (\ref{einstein}) 
on the brane as well as the above equation one deduces that the effective brane energy momentum tensor on the brane is conserved with respect to the induced metric. 
From this and equations 
(\ref{SRS}) and (\ref{SDGP}) (as well as from Bianchi identities in the latter case) 
 one deduces that the brane matter energy momentum tensor is also conserved with respect to the induced metric
 in both models (I) and (II).
One has then  for the RS and DGP models  
\beq 
D^\mu S_{ \mu \nu} = 0,  \label{consS} \\
D^\mu  T^{(M)}_{ \mu \nu} = 0.  \label{consmat}
\eeq
One also deduces from equation (\ref{effein}) and Bianchi identities that 
\beq \label{consweyl}
D^\mu {\cal E}_{\mu \nu} = \kcq D^\mu \Pi_{\mu \nu},
\eeq
which can be considered as a conservation  equation for the Weyl's fluid. In general the energy momentum tensor of the latter is indeed not conserved with respect to the induced metric, since the right hand side of the above equation does not vanish,  as will be seen below, we will however still refer to the equations deduced from (\ref{consweyl}) as conservation equations for the Weyl's fluid.

\subsection{Cosmological background} \label{COSBACK}
To deal with the background cosmological evolution, it is convenient to use a GN coordinate system (\ref{GN}) and to consider the following ansatz for the metric
\beq \label{backmet}
ds^2_{(5)} = -n^2(t,y) dt^2 + a^2(t,y) \delta_{ij} dx^i dx^j + dy^2, 
\eeq
where the 3D metric $\delta_{ij}$ is a flat euclidean metric (we will only consider in this paper the case of a spatially flat universe). One can further choose a time parametrization such that the function $n$ is set to $1$ on the brane ($n_{(b)} =1$), in which case  the induced metric (\ref{induite}) is simply given by 
\beq \nonumber
ds^2_{(4)} = -dt^2 + a^2_{(b)} \delta_{ij} dx^i dx^j
\eeq
(where $a_{(b)} \equiv   a(t,y=0)$) and 
is of FLRW form. Considering comoving observers to sit at fixed comoving coordinates $x^i$ on the brane, 
$t$ is then simply the cosmological time on the brane. 
Accordingly with the symmetries of (\ref{backmet}) the effective energy momentum tensor of the brane is taken of the form 
\beq \nonumber
S^{\mu}_{\nu} = \delta(y) \mbox{diag} \left(-\rho,P,P,P \right).
\eeq
We point out that 
in the above expression $\rho$ and $P$ are not to be understood as {\it real matter} energy density and pressure, 
that will be denoted $\rho_\MM$ and  $P_\MM$, but
as their {\it effective matter} counterparts.
 Their relation to real matter energy density and pressure is model dependent.   
For example in the case of RS model  one has 
\beq \nonumber
\rho = \rho_\MM + \Lq.
\eeq
The expressions for $\rho$ and $P$ as functions of $\rho_\MM$ and $P_\MM$ in the RS and DGP models are given in appendix \ref{appA}. 
The junction conditions (\ref{Israel}) are given for the background by 
\beq \label{backjun}
\frac{\abp}{\ab} &=& -\frac{\kcd}{6} \rho  \\
\frac{\nbp}{\nb} &=& \frac{\kcd}{6}(3P + 2 \rho)  \label{junn} 
\eeq
where a prime  denotes a derivation with respect to $y$. 
The behavior of the scale factor on the brane can be obtained solving the equations
(\ref{effein}) and (\ref{consS}) which for the background reduce to the two independent equations 
valid for a bulk with a vanishing background Weyl's tensor (we will always assume this to be the case in the rest of this paper) 
\beq
H^2 &=& \frac{\Lc}{6} + \frac{\kcq}{36} \rho^2, \label{Brafried} \\
\label{consback}
\dot{\rho} &=& - 3 H (P+\rho),
\eeq
where
a dot means a derivation with respect to time $t$,  and 
 $H$ is the Hubble parameter on the brane given by $H \equiv \abd / \ab$.

One finds  
in the Randall-Sundrum model, using the above equations,  a transition \cite{Cline:1999ts,Binetruy:2000hy} from an early non standard phase to a late time standard cosmological phase which dynamics is given at leading order by usual Friedmann's equations
\beq
H^2 &=& \frac{\kqd}{3} \rho_\MM,  \nonumber \\
\dot{\rho}_\MM &=& - 3H (P_\MM + \rho_\MM). \label{StanFRIE}
\eeq 
The transition happens approximately when the Hubble radius $H^{-1}$ becomes larger than the threshold $r_{(5)}$. 

For the DGP model,
one has a transition from an early standard cosmological phase, governed by equations (\ref{StanFRIE}), 
to a late time non standard phase \cite{Deffayet:2001uy}. The transition happens approximately when 
 $H^{-1}$ crosses the length scale $r_c$. A conservative bound on $r_c$ (in agreement with other bounds, see \cite{Dvali:2001gx,Gruzinov:2001hp})
is then given by demanding  that $r_c$ should be  larger than (or  of the order of) today's Hubble radius. As a consequence the parameter $\Upsilon$ defined by 
\beq \label{Upsi}
\Upsilon = \frac{1}{H r_c}, 
\eeq
is always smaller than (or of the order of) one in the past. 

An explicit expression of the bulk metric can also be obtained in the coordinate system 
(\ref{backmet}) \cite{Binetruy:2000hy}, we quote in appendix \ref{appA0} some useful identities concerning the solutions for the background
metric both in the bulk and on the brane.

\section{Scalar metric perturbations in the bulk} \label{METPER}In this section, we first  (subsection \ref{DEFMET})
introduce our notations (mostly the same as in  \cite{Bridgman:2001mc}) 
for scalar metric perturbations and gauge invariant variables.
We then remind, in subsection  \ref{MASTER}, 
 the expression of Mukohyama's master equation
\cite{Mukohyama:2000ui}
 as well as the relation between the master variable and gauge invariant variables.  Eventually we discuss there
some properties of the Master equation in relation with issues related to initial conditions in the bulk and boundary conditions on the brane for the DGP and RS models.

\subsection{Metric perturbations  and gauge invariant variables} \label{DEFMET}
The five dimensional perturbed metric element is taken to be
\beq \label{linepert}
g_{AB} = \left(\begin{array}{ccc}
-n^2(1+2A) & a^2 B_{|i} & n A_y \\
a^2B_{|i} & a^2\left[(1+2 {\cal R})\delta_{ij}+ 2 E_{|ij}\right]&a^2B_{y|i}\\
n A_y & a^2B_{y|i} &1+2A_{yy}
\end{array}\right),
\eeq
where we have kept only scalar perturbations, and $_{|i}$ denotes
 a differentiation with respect to the comoving coordinate $x^i$.
A scalar coordinate transformation is defined as
\beq \label{gaugetrans} \label{5Dt}
t &\rightarrow& t + \delta t, \\
x^i &\rightarrow& x^i + \delta^{ik} \delta x_{|k} ,\label{5Dx}\\
y &\rightarrow& y + \delta y, \label{5Dy}
\eeq
\label{PCGI-ART1}
and induces the following transformation on the scalar quantities
\beq \nonumber
A &\rightarrow& A -\dot{\delta t} - \frac{\dot{n}}{n} \delta t -
\frac{n^\prime}{n} \delta y, \\ \nonumber
{\cal R}&\rightarrow & {\cal R} - \frac{\dot{a}}{a} \delta t -
\frac{a^\prime}{a} \delta y, \\ \nonumber
B&\rightarrow &  B + \frac{n^2}{a^2} \delta t - \dot{\delta x},\\ \nonumber
B_y &\rightarrow & B_y - \delta x^\prime - \frac{1}{a^2} \delta y, \\ \nonumber
E&\rightarrow &  E - \delta x,\\ \nonumber
A_y &\rightarrow & A_y + n \delta t^\prime - \frac{1}{n} \dot{\delta y}, \\ \nonumber
A_{yy}&\rightarrow & A_{yy} - \delta y^\prime.
\eeq
In the following a gauge transformation involving a zero $\delta y$ will be denoted as a {\it 
4D-gauge transformation} as opposed to a gauge transformation involving a non zero $\delta y$ which will be referred to as a  {\it 
5D-gauge transformation}.
For further reference we  note that 
 the (perturbed) brane position can be defined locally by $y=\xi(x^i,t)$.
 \label{GIV}
We will also use the 3D spatially gauge invariant variables \cite{Bridgman:2001mc}
\beq
\sigma &\equiv& -B + \dot{E}, \nonumber \\
\sigma_y & \equiv& -B_y +E^\prime,  \nonumber
\eeq
transforming as
\beq
\sigma &\rightarrow &\sigma - \frac{n^2}{a^2} \delta t  \nonumber\\
\sigma_y &\rightarrow &\sigma_y + \frac{1}{a^2} \delta y  \nonumber
\eeq
under the coordinate transformation (\ref{gaugetrans}).
We also define the variable $\Psi$ and $\Phi$ invariant under 4D-gauge transformations by 
\cite{Mukhanov:1992me}
\beq
\Phi &=& A - \frac{1}{n} \left(\frac{a^2 \sigma }{n} \right)^.,  \nonumber\\
-\Psi &=& {\cal R} - \frac{\dot{a}a}{n^2} \sigma.  \nonumber
\eeq 
From equations (\ref{5Dt}-\ref{5Dy}), one sees that the most general (scalar) 5D gauge transformation involves 3  unknown functions, implying that one can define out of the 7 perturbation variables appearing in (\ref{linepert}), four independent variables invariant under 
a 5D-gauge transformation \cite{Mukohyama:2000ui,cvdb}.
We take the latter to be given by  $\tilde{A}$, $\tilde{A}_y$, $\tilde{A}_{yy}, \tilde{R}$, expressed as 
\cite{Bridgman:2001mc}
\beq
\tilde{A} &=& A - \frac{1}{n}\left(\frac{a^2 \sigma}{n}\right)^. + \frac{n^\prime}{n} a^2 \sigma_y,  \nonumber\\ \nonumber
\tilde{A}_y &=& A_y + \frac{(a^2 \sigma_y)^.}{n}+\frac{(a^2 \sigma)^\prime}{n} - 2 \frac{n^\prime}{n^2}a^2 \sigma, \\ \nonumber
\tilde{A}_{yy} &=& A_{yy} + (a^2 \sigma_y)^\prime, \\ \nonumber
\tilde{\cal R} &=& {\cal R} + a a^\prime \sigma_y - \frac{a \dot{a}}{n^2} \sigma. 
\eeq

\subsection{Master equation} \label{MASTER}
As  shown by Mukohyama \cite{Mukohyama:2000ui},  the linearized scalar  Einstein's equations in a
  maximally symmetric bulk can be conveniently
solved introducing a master variable $\Omega$
 which obeys a PDE in the bulk, the master equation.
The latter, when $\Omega$ has a non trivial  dependence  in the comoving coordinates $x^i$, reads in a  GN coordinate system (\ref{backmet}) 
\beq 
\label{FQ1}
\left(\frac{\dot{\Omega}}{na^3}\right)^. + \left(\frac{\Lc}{6}-
\frac{\Delta}{a^2}\right) \frac{n \Omega}{a^3} - \left(\frac{n
\Omega^\prime}{a^3} \right)^\prime=0,
\eeq
where $\Delta$ is defined by $\Delta = \delta^{ij} \partial_j \partial_i$. 
In the rest of this article, we will implicitly consider all the perturbations as Fourier transformed  
with respect to the $x^i$s, in order to do a mode by mode analysis.
In particular  (\ref{FQ1}) can be rewritten as 
\beq \nonumber
{\cal D}_\Delta \Omega =0, 
\eeq
where ${\cal D}_\Delta$ is a second order hyperbolic differential operator acting on $y$ and $t$ dependent functions (in the GN system), and  $\Delta$ 
is understood to be replaced by  $-\vec{k}^2$, where $\vec{k}$ is the comoving momentum. 
 Equation (\ref{FQ1}) is then only valid when $\vec{k}^2$ does not vanish \cite{Mukohyama:2000ui}, which is the only case of interest as far as cosmological perturbations are concerned. 
The gauge invariant metric perturbations are then related to $\Omega$ by \cite{Mukohyama:2000ui,Bridgman:2001mc}
\beq \label{EQ1}
\tilde{A} &=& -\frac{1}{6a} \left(2 \Omega^{\prime \prime} + \frac{1}{n^2}
 \ddot{\Omega} + \frac{\Lc }{6} \Omega -\frac{\dot{n}}{n^3} \dot{\Omega} - \frac{n^\prime}{n} \Omega^\prime \right),
\\ \label{EQ6}
\tilde{A}_y &=& \frac{1}{an} \left(\dot{\Omega}^{\prime} - \frac{n^\prime}{n} \dot{\Omega} \right),
\\ \label{EQ4}
 \tilde{A}_{yy} &=& \frac{1}{6a} \left( \Omega^{\prime \prime} +
\frac{2}{n^2} \ddot{\Omega} - \frac{\Lc}{6} \Omega  - 2 \frac{\dot{n}}{n^3} \dot{\Omega} - 2 \frac{n^\prime}{n} \Omega^\prime \right),
\\ \label{EQ20}
\tilde{{\cal R}} &=& \frac{1}{6a} \left( \Omega^{\prime \prime} - \frac{1}{n^2} \ddot{\Omega} + \frac{\Lc}{3} \Omega + \frac{\dot{n}}{n^3} \dot{\Omega} + \frac{n^\prime}{n} \Omega^\prime \right).
\eeq
The master equation enables a priory to solve for the evolution of brane world cosmological perturbations, once initial conditions in the bulk  and a suitable boundary condition on the brane are provided, as we now discuss.
\label{causmast}

\begin{figure}
\centering
\psfrag{C}{ brane cosmic time}
\psfrag{T}{ }
\psfrag{A}{A}
\psfrag{G}{D}
\psfrag{H}{C}
\psfrag{t}{X}
\psfrag{E}{B}
\psfrag{F}{X=0}
\psfrag{x}{T}
\psfrag{i}{Initial time}
\psfrag{g}{Big Bang}
\psfrag{b}{brane}
\psfrag{I}{E}
\psfrag{J}{F}
\resizebox{14cm}{8cm}{\includegraphics{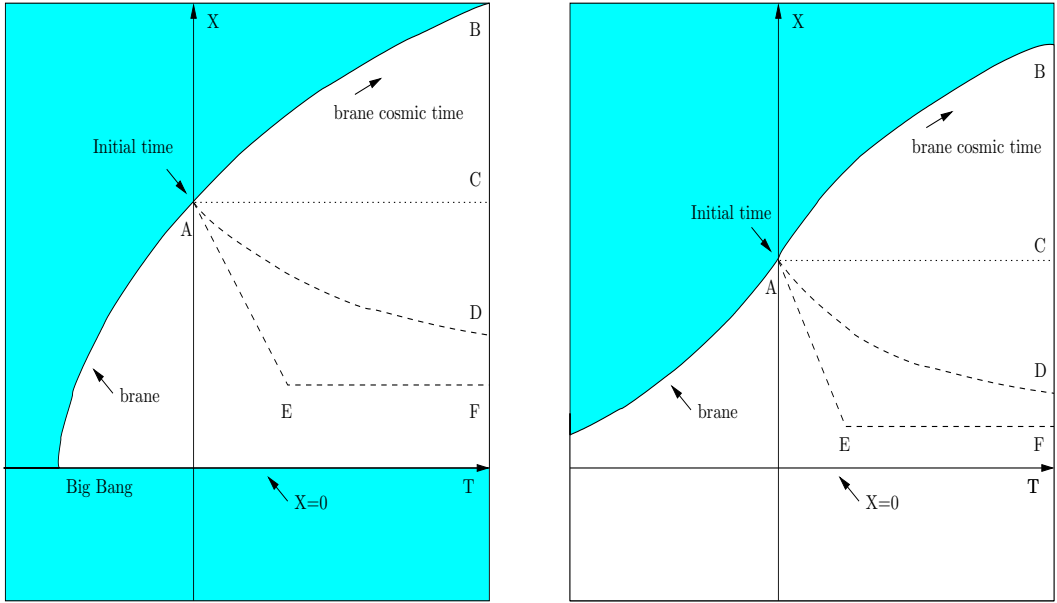}}
\caption{Schematic representation of the bulk space-time with the
  brane trajectory in the characteristic $X,T$ coordinates (of equation (\ref{MINKOCOS})) for the DGP model.
The left figure is the case with a Big Bang (with e.g. a radiation dominated universe), while the right one corresponds to a case with inflation. The $X=0$ line is a coordinate singularity. The gray (cyan) part is cutoff and the complete bulk space-time is made out of the remaining part, glued to a copy of itself along the brane. If a given initial instant is chosen along the brane from which one wishes to evolve cosmological perturbations, 
one needs to specify a boundary condition along the brane (AB), and
initial data in the bulk (e.g along (AD)) that may be in part
characteristic data (as along (AEF)).
The future of the brane events with cosmic time larger than initial time lies above the dotted line (AC).}
\label{Fig1}
\end{figure}

Let us first deal with the DGP model, where the bulk is simply a slice of $5D$ Minkowski.
In this case, the metric (\ref{backmet}) can be put in the form 
\label{PCGI12p26}
\beq \label{MINKOCOS}
ds^2 = -dX dT + X^2 dx^i dx^j \delta_{ij}
\eeq
\label{PCGI22p18}
where $X$ is given by $a(t,y)$ and $T$ is a function of the Gaussian normal coordinates 
$y$ and $t$ which can be easily obtained e.g. by the coordinate change given in \cite{Deruelle:2000ge}. 
The background trajectory of the brane $X_\bb(t)$ and $T_\bb(t)$ is then given by  
\label{GSRES}
\beq
X_\bb &=& a_\bb,  \nonumber\\
\dot{T}_\bb &=& {\dot{a}^{-1}_\bb}, \nonumber
\eeq 
with $a_\bb$ obtained by solving equations (\ref{Brafried}) and (\ref{consback}).
In the coordinate system $(X,T)$, equation (\ref{FQ1}) reads
\beq
{\cal D}_\Delta \equiv -  \partial_{X} \partial_T  \Omega + \frac{3 \partial_T \Omega}{2 X} + \frac{\Delta \Omega}{4 X^2} =0. \label{nullDGP}
\eeq
where the hyperbolic nature of ${\cal D}_\Delta$ is manifest. The characteristics 
of ${\cal D}_\Delta$ are simply the null lines $X=constant$, $T=constant$. 
If one assumes that $\dot{a}_\bb$ is always positive, then $X_\bb$ is an increasing function of time, in which case  the trajectory of the brane can be schematized as shown on figure \ref{Fig1}. The brane background  trajectory is then non characteristic.
Let us now assume that a given initial time is chosen along the brane, from which one wishes to evolve cosmological perturbations.  One then needs 
to specify initial data in the bulk , e.g. as shown in figure  \ref{Fig1}, as well as a proper boundary condition along the brane. 
We will not discuss here how the bulk initial conditions are
provided\footnote{The latter can be thought of as being given by some
  early universe physics.}; we 
rather aim, as far as the issue of initial and boundary data are concerned, at clarifying the expression of the boundary condition on the brane.
The latter can be imposed along the {\it background} brane trajectory,
and will be given and discussed  more explicitly in section
\ref{bound}\footnote{We also discuss more in appendix \ref{appDD} the
  issue of bulk initial data in relation with the boundary condition
  found in this section.}.
We further note here  that the bulk initial data 
 may involve in general characteristics initial data\footnote{as opposed to Cauchy type of data specified on non characteristic curves. Such characteristics initial data can also be considered as a model dependent boundary condition in the bulk.}  (meaning here data specified on a characteristic $X= constant$ line). 
Eventually, we stress that the  
 coordinate system $(X,T,x^i)$ is singular in $X=0$. One can further transform  the metric 
(\ref{MINKOCOS}) to a flat Minkowski metric of the form 
\beq \label{canoMIN}
ds^2 = -4 dY^+dY^- + (dY^1)^2 + (dY^2)^2 + (dY^3)^2,
\eeq
where $Y^+$ and $Y^-$ are light cone coordinates and one has $X= Y^-$. In this coordinate system, the master equation reads  
\beq  \nonumber
{\cal D}_\Delta \equiv \eta^{AB}\partial_A \partial_B \Omega + \frac{3 \partial_{+} \Omega }{Y^-} =0, 
\eeq
where $\eta_{AB}$ is the metric appearing in (\ref{canoMIN}). 
Using this coordinate system, one can easily see that a brane with a Big Bang is 
a conoidal hyper-surface with  a vertex  that can be taken to be the origin. The brane is then inside the light cone of the origin which represents the Big Bang singularity. The latter also extends along the null line $Y^-=0$ on the light cone of the origin, to which the brane is tangent
 in strong analogy to what is happening when the bulk is  a slice of AdS$_5$ \cite{Ishihara:2001qe}.
 One notes that the master equation is singular along this line, where
 the scalar-vector-tensor splitting of Minkowski$_5$ gravitational
 perturbations becomes ill-defined\footnote{This is because this
   splitting is defined by a particular choice of coordinates on
   Minkowski$_5$, e.g defined by equation (\ref{MINKOCOS}), which
   breaks down along the $Y^-\equiv X =0$ line.}. This is not without
 importance if one wants to set initial data close to the initial singularity.

In the case of the RS model, the bulk is a slice of $AdS_5$ space-time, and 
the metric (\ref{backmet}) can be put in the form 
\label{PCGI31p10-15}
\beq \nonumber
ds^2 = \frac{r^2_{(5)}}{z^2} \left( - d\tau^2 + dz^2 + \eta_{ij} dx^i dx^j \right), 
\eeq
where one has $z = r_{(5)} / a(t,y)$ and  $\tau$ is a function of the GN coordinates $y$ and $t$ that can be found using the work of \cite{Mukohyama:1999wi}.
The brane trajectory is given by \cite{Kraus:1999it}
\beq \nonumber
z_\bb &=& \frac{r_{(5)}}{a_\bb}, \\ \nonumber
\dot{\tau}_\bb &=& \frac{1}{a_\bb} \sqrt{1+ r^2_{(5)} \frac{ \dot{a}_\bb^2}{a^2_\bb} }.
\eeq
In the coordinate system $(z,\tau)$, equation (\ref{FQ1}) reads \cite{Mukoh2} 
\beq \nonumber
{\cal D}_\Delta \equiv  \partial^2_z \Omega - \partial^2_\tau \Omega  + \frac{3 \partial_z \Omega}{z} + \frac{\Omega}{z^2} + \Delta \Omega =0.
\eeq
Using the null coordinates $X\equiv \tau+z $ and $T= \tau-z$, this can be further rewritten
\beq
{\cal D}_\Delta \equiv - \partial_X \partial_T \Omega + \frac{3(\partial_X-\partial_T) \Omega}{X-T} + \frac{4 \Omega}{(X-T)^2} + \Delta \Omega = 0. \label{nullRS}
\eeq
We also see clearly the hyperbolic nature of ${\cal D}_\Delta$. 
The characteristics curves of the latter are the null lines $X = constant$ and $T=constant$, and the brane trajectory is also non characteristic. 
We will not discuss here in more details this case since it is strongly analogous to the previous one. 

\section{Matter and Weyl perturbations on the brane} \label{MATPER}
In order to obtain a boundary condition on the brane and to recover
perturbed effective Einstein's equations from the master equation, one
needs relations between matter, Weyl's fluid,  and the master
variable. These can be obtained by using 
limiting values on the brane of derivatives of bulk metric perturbations which are related to effective matter and Weyl's fluid perturbations
 (that we define in the next subsection) through Israel's junction conditions (\ref{Israel}) and the Weyl's fluid definitions (\ref{Weyldef}-\ref{defTW}). The corresponding expression,  in a convenient gauge defined in 
subsection \ref{GAUCHOI},
are reminded in subsection \ref{JUNCON}.

\subsection{Perturbations of effective matter,  real matter and Weyl's fluid energy-momentum tensor on the brane}
Following standard cosmological perturbation theory,  we can always decompose 
the scalar part of the perturbations of the brane effective energy momentum tensor as 
\beq
\delta S^0_0 &=& -\delta \rho, \label {dS00}\\ 
\delta S^0_i &=& \delta q_{|i},  \label{dS0i} \\
\delta S^i_j &=& \delta P \delta^i_j +  \left(\Delta^i_j - 
\frac{1}{3} \delta^i_j \Delta  \right) \delta \pi, \label{dSij}  
\eeq
where $ \Delta^i_j$ is defined by 
\beq \nonumber
 \Delta^i_j = \delta^{ik} \partial_k \partial_j, 
\eeq
so that one has $\Delta = \Delta^i_i$.
Similarly the scalar real matter perturbations are defined from perturbations of the real matter energy momentum tensor
$T_{\mu \MM}^\nu$ and read 
\beq
\delta {T}^0_{0 \MM} &=& -\delta {\rho}_\MM, \label {dTM00}\\
\delta {T}^0_{i \MM} &=& \delta {q}_{ \MM |i}, \label{dTM0i}\\
\delta {T}^i_{j \MM} &=& \delta {P}_\MM \delta^i_j +  \left(\Delta^i_j - 
\frac{1}{3} \delta^i_j \Delta  \right)   \delta {\pi}_{ \MM } \label{dTMij}.  
\eeq
The relation between the effective matter perturbations (\ref{dS00}-\ref{dSij}) and the {\it real} matter perturbations 
(\ref{dTM00}-\ref{dTMij}) in the two different models I/ and II/ mentioned in the introduction can easily be read off from equations (\ref{SRS}) and (\ref{SDGP}) respectively. They will be given in section \ref{bound}.  
Similarly, 
 the Weyl's fluid scalar 
perturbations can be decomposed as \cite{Roy1,Langlois:2001ph}
\beq \label{dE00} 
\delta {\cal E}^0_0 &=& \kqd \delta \rho_{\EE}, \\ \label{dE0j}
\delta {\cal E}^0_j &=& -\kqd \delta q_{\EE |j}, \\ \label{dEij}
\delta {\cal E}^i_j &=& 
-\kqd \left(\delta P_{\EE} \delta^i_j +  \left(\Delta^i_j - \frac{1}{3} \delta^i_j \Delta \right) \delta \pi_{\EE} \right).
\eeq
From equation (\ref{traceW}) one can deduce that the Weyl's fluid perturbation obeys the 'equation of state' of radiation, namely one has 
\beq \label{Weyleq}
\delta P_\EE = \frac{1}{3} \delta \rho_\EE.
\eeq

\subsection{Gauge choices} \label{GAUCHOI}
 From equation (\ref{gaugetrans}), one sees easily that one can simultaneously
choose a gauge where $A_{yy}$, $A_{y}$ and $B_{y}$ are vanishing and the brane
sits at $y=0$ \cite{Langlois:2001ph}. This gauge choice put the perturbed metric in the GN form 
(\ref{GN}). This however does not completely fix the gauge, but
one can further make $y-independent$ shifts of $x^i$ and $t$ without
spoiling the gauge conditions enumerated above; this gauge freedom is simply the
usual 4D gauge freedom, and one can use is to go to the usual (4D) longitudinal
gauge on the brane. We will refer to such a gauge choice as 
a Gaussian Normal Longitudinal gauge (denoted as a GNL gauge) in the following. 
To summarize, the GNL gauge is defined by the conditions
\beq
\bar{A}_y &=& 0, \\
\bar{B}_y &=& 0, \\
\bar{A}_{yy} &=& 0, \\
\bar{\xi} &=& 0, \\
\bar{\sigma}_\bb &=& 0, \label{sigb}
\eeq
where the last equality defines the 4D longitudinal gauge on the brane, and we have used bars to differentiate quantities in the GNL gauge from quantities in arbitrary gauge (e.g. $\bar{A}$ vs. $A$). 
In the rest of this paper, we will always work in the GNL gauge
 when dealing with quantities of which we take limiting values on the brane, and in order to alleviate the notations we will drop the bars. 
In the GNL gauge, one has the following identities valid in the bulk (say in the vicinity of the brane)
\beq
\sigma_y &=& E^\prime, \\
\sigma &=& -B + \dot{E}, \\
\tilde{A} &=& A - \frac{1}{n}\left(\frac{a^2 \sigma}{n}\right)^. + \frac{n^\prime}{n} a^2 E^\prime \label{A}, \\
\tilde{A}_y &=& \frac{(a^2 E^\prime)^.}{n}+\frac{(a^2 \sigma)^\prime}{n} - 2 \frac{n^\prime}{n^2}a^2 \sigma \label{Ay},\\
\tilde{A}_{yy} &=&(a^2 E^\prime)^\prime \label{Ayy},\\
\tilde{\cal R} &=& {\cal R} + a a^\prime E^\prime - \frac{a \dot{a}}{n^2} \sigma, \label{R}
\eeq
and on the brane  
\beq \nonumber
\left( A \right.&=& \left.\Phi\right)_\bb, \\ \nonumber
\left( {\cal R} \right. &=&\left. - \Psi\right)_\bb, \\ \nonumber
\left( \tilde{A} \right. &=&\left.  \Phi + \frac{n^\prime}{n} a^2 E^\prime\right)_\bb ,\\ \nonumber
\left( \tilde{A}_{y} \right. &=& \left.\frac{(a^2 E^\prime)^\cdot}{n} + \frac{a^2}{n} \left( -B^\prime + \dot{E}^\prime \right) \right)_\bb ,\\ \nonumber
\left( \tilde{A}_{yy} \right. &=&\left. a^2 E^{\prime \prime} + 2 a^\prime a E^\prime\right)_\bb, \\ \nonumber
\left( \tilde{\cal R} \right. &=&\left. -\Psi + a a^\prime E^\prime\right)_\bb. 
\eeq
We  underline that the gauge condition (\ref{sigb}) does not imply the vanishing of  $y-$derivatives of $\sigma$ on the brane, that have been kept above.

\subsection{Junction conditions for effective matter and Weyl's fluid} \label{JUNCON}

The Israel junction conditions (\ref{Israel})  in the GNL gauge lead
to the following expressions for the $y-$derivatives of the metric perturbations on the brane   \cite{Bridgman:2001mc}
\beq
{A}^\prime_\bb &=& \frac{\kcd}{6}\left(3 \delta P + 2 \delta \rho\right) \label{DPdef}, \\
{{\cal R}}^\prime_\bb & =& \frac{1}{6} \kcd \left( \Delta \delta \pi - \delta \rho \right), \label{Rpdef}\\
{B}^\prime_\bb &=& \kcd \frac{n_\bb^2}{a_\bb^2} \delta q \label{Dddef},\\
{E}^\prime_\bb &=& -\frac{1}{2} \kcd \delta \pi . \label{Ep}
\eeq
One has also \label{[PCGI 14, p16]}
\beq \nonumber
{\sigma}^\prime_\bb &=& -\kcd \frac{n_\bb^2}{a_\bb^2} \delta q - \frac{1}{2} \kcd \delta \dot{\pi}.
\eeq
Similarly, one can get expressions relating the second $y-$derivatives on the brane of metric perturbations to 
Weyl's fluids  (and effective matter) perturbations. They read in the GNL gauge \cite{Bridgman:2001mc} (those expressions are obtained 
using   the expression of the Weyl's tensor 
perturbations (\ref{dE00}-\ref{dEij}), the definition (\ref{Weyldef}) 
and some bulk perturbed Einstein's equations)
\label{A'':PCGI Ann 13}
\beq \label{DPEdef}
\kqd \left( \delta P_{\EE} + \frac{2}{3} \delta \rho_{\EE} \right) &=& - \left\{ {{A}}^{\prime \prime} + 2 \frac{n^\prime}{n} {A}^\prime\right\}_\bb, \\ \label{DQEdef}
\kqd \delta q_{\EE} &=& -\frac{1}{2} \frac{a_\bb^2}{n_\bb^2} \left\{ {B}^{\prime \prime} +
 \left(3\frac{a^\prime}{a} - \frac{n^\prime}{n} \right) {B}^\prime\right\}_\bb, \\ \label{DPIEdef}
\kqd \delta \pi_{\EE} &=& \left\{ E^{\prime \prime} + 2 \frac{a^\prime}{a} E^\prime\right\}_\bb, \\ \label{DREdef}
\kqd \delta \rho_{\EE} &=&\left\{ 3 \left( {R}^{\prime \prime} + 2 \frac{a^\prime}{a} {R}^\prime\right) + \Delta E^{\prime \prime} + 2 \frac{a^\prime}{a} \Delta E^\prime \right\}_\bb.
\eeq
One can consider the above expression  (\ref{DPdef}-\ref{Ep}) and   (\ref{DPEdef}-\ref{DREdef})  as 
definitions 
of  effective matter energy momentum perturbations (e.g. $\delta q$) or Weyl's fluid perturbations  (e.g. $\delta q_\EE$) 
respectively in terms of first (e.g. $B^\prime$) 
and second derivatives (e.g. $B^{\prime \prime}$)
transverse to the brane of bulk metric perturbations. There is  indeed a  one to one correspondence between the former and the latter, and we will in the following call all equations  (\ref{DPdef}-\ref{DREdef}) junction conditions (respectively for effective matter or Weyl's fluid perturbations).
Following this line, we have not simplified   
equation  (\ref{DPEdef}) using the  equation of state (\ref{Weyleq}), to keep the correspondence manifest. We will do so until the end of next section.

\section{Linearized  effective Einstein's and conservation equations on the brane from the master equation}
\label{PertEin}
With obvious notations, the perturbed 4D effective Einstein's equations (\ref{effein}) reads 
\label{(PCGI 1,p 20)} 
\beq \label{Perteffein}
\delta G^{(4)\mu}_\nu = -\delta {\cal E}^\mu_\nu + \kcq \delta \Pi^\mu_\nu, 
\eeq
where one can insert the definitions (\ref{dS00}-\ref{dSij}) and (\ref{dE00}-\ref{dEij}) given in the previous section to get 
expressions relating the perturbed induced metric on the brane to effective matter and Weyl's fluid perturbations.  Those expressions  can then be easily compared to usual 4D cosmological perturbation equations \cite{Langlois:2001ph}. The latter are given in appendix \ref{4Dpert}. The components of the  linearized effective Einstein's equations are gathered in appendix \ref{B2}  together with the ones of linearized conservation  equations (\ref{consS}) and (\ref{consweyl}). 

We would like here to show how those last equations can be recovered  from the Master equation (\ref{FQ1}) and the junction conditions for effective matter and Weyl's fluid given  in previous subsection.
The outline of our derivation is to built 'constraints' valid everywhere in the bulk on the gauge invariant variables 
 $\tilde{A}, \tilde{A}_y, \tilde{A}_{yy}$ and $\tilde{R}$ (as well as on their $y-$ and $t-$ derivatives) out of the 
master equations and the definitions (\ref{EQ1}-\ref{EQ20}). The latter constraints can be thought of as reconstruction of linearized 5D Einstein's equations in the bulk out of the master equation, in confirmation of Mukohyama's work \cite{Mukohyama:2000ui}. We then take limiting values on the brane of those constraints, where we replace the gauge invariant variables by their expressions as functions of effective matter, Weyl's fluid and induced metric that can be obtained through the junction conditions. Those will yield the sought for equations. Let us now proceed.

First, using equations (\ref{A}-\ref{R}), the derivatives with respect to $y$ of those equations, the  junction conditions for perturbations
(given in subsection \ref{JUNCON}) 
and background (equations (\ref{backjun}-\ref{junn})), as well as the equation of motion for the background (gathered  in appendix \ref{appA}),  we get 
  the following expressions 
relating limiting values on the brane of gauge invariant variables (and their $y-$derivatives) to effective matter, Weyl's fluid and induced metric perturbations
\label{ PCGI ANN10 et 1}
 \beq
\label{Atb}
\tilde{A}_\bb &=&\left\{ \Phi -  \frac{\kcq}{12} a^2 
\left( 3P + 2 \rho\right) \delta \pi
\right\}_\bb,\\
\label{Atyb}
\tilde{A}_{y \bb} &=& \left\{- \kcd \delta q - \kcd a \left( a \delta \pi \right)^\cdot \right\}_\bb
,\\
\label{Atyyb}
\tilde{A}_{yy \bb} &=& a^2_\bb \kqd \delta \pi_\EE, \\
\label{Rtb}
\tilde{\cal R}_\bb &=& -\Psi_\bb + \frac{\kcq}{12}  a^2_\bb \rho \delta \pi, \\
\tilde{A}^\prime_\bb &=& \left\{\frac{\kcd}{6}\left(3 \delta P + 2 \delta \rho \right) + \kcd \delta \dot{q} + \frac{\kcd}{2} a \left( a \delta \pi \right)^{\cdot \cdot} \nonumber \right.  \\
&&\left. + \frac{\kcd}{6} a^2 ( 3P + 2 \rho )\left[ \kqd \delta \pi_\EE + \frac{\kcq}{4}(P+\rho)  \delta \pi \right] \right\}_\bb,
 \label{Atbp}\\
\tilde{A}^\prime_{y\bb} &=&\left\{ \frac{\kcq}{6}  (5\rho+ 6 P) \left[ a \left( a \delta \pi \right)^\cdot + \delta q \right] \nonumber \right.\\
&&\left.+ 2 a \kqd \left( a \delta \pi_\EE \right)^\cdot + 2 \kqd \delta q_\EE \right\}_\bb\label{Atybp}, \\
\tilde{\cal R}^\prime_\bb &=& \left\{ -\frac{\kcd}{6} \delta \rho + \frac{\kcd}{6} \Delta \delta \pi + 
\frac{\kcd}{2} \dot{a} \left( a \delta \pi \right)^\cdot + \frac{\dot{a}}{a} \kcd \delta q \nonumber \right. \\
&&\left.- \frac{\kcd}{6} \kqd \rho a^2 \delta \pi_\EE \right\}_\bb
\label{Rtbp}.
\eeq
The value on the brane of $\tilde{A}^\prime_{yy}$ can however not be computed at this stage
as can be seen from equation (\ref{Ayy}). It would indeed require to know  $E^{\prime \prime \prime}_\bb$ which is not given by the junction conditions.   Let us now derive three constraints on the  gauge invariant variables $\tilde{A}, \tilde{A}_y, \tilde{A}_{yy}, \tilde{R}$ and their first derivatives  $\tilde{A}^\prime , \tilde{A}^\prime_y, \tilde{R}^\prime$, which limiting value on the brane we know from the above expressions.

Adding equations (\ref{EQ1}), (\ref{EQ4}) and (\ref{EQ20}) one first gets  
that everywhere in the bulk  \cite{cvdb,Bridgman:2001mc}
\beq \label{GQ1} 
\tilde{A} + \tilde{A}_{yy} + \tilde{\cal R} = 0,
\eeq
which is the first constraint we will use latter.

We now consider the five equations obtained taking  first derivatives of equations (\ref{EQ1}), (\ref{EQ6}), (\ref{EQ4}) with respect to $y$ and $t$. They read symbolically  
\beq
\tilde{A}^{\cdot} &=& -\left\{\frac{1}{6a} \left(2 \Omega^{\prime \prime} + \frac{1}{n^2}
 \ddot{\Omega} + \frac{\Lc }{6} \Omega -\frac{\dot{n}}{n^3} \dot{\Omega} - \frac{n^\prime}{n} \Omega^\prime \right) \right\}^{\cdot}, \\
\tilde{A}^\prime &=& -\left\{\frac{1}{6a} \left(2 \Omega^{\prime \prime} + \frac{1}{n^2}
 \ddot{\Omega} + \frac{\Lc }{6} \Omega -\frac{\dot{n}}{n^3} \dot{\Omega} - \frac{n^\prime}{n} \Omega^\prime \right) \right\}^{\prime}, \\
\tilde{A}_y^{\cdot} &=&\left\{\frac{1}{an} \left(\dot{\Omega}^{\prime} - \frac{n^\prime}{n} \dot{\Omega} \right) \right\}^{\cdot}, \\
\tilde{A}_y^{\prime} &=&\left\{\frac{1}{an} \left(\dot{\Omega}^{\prime} - \frac{n^\prime}{n} \dot{\Omega} \right) \right\}^{\prime}, \\
\tilde{A}_{yy}^{\cdot} &=& \left\{\frac{1}{6a} \left( \Omega^{\prime \prime} +
\frac{2}{n^2} \ddot{\Omega} - \frac{\Lc}{6} \Omega  - 2 \frac{\dot{n}}{n^3} \dot{\Omega} - 2 \frac{n^\prime}{n} \Omega^\prime \right)  \right\}^{\cdot}
\label{EQ5}.
\eeq
 These five equations, together with 
equations  (\ref{EQ1}), (\ref{EQ6}), (\ref{EQ4}), the master equation (\ref{FQ1})
and its derivative with respect to $y$ and $t$ 
can be considered as forming a system $({\cal S}_0)$ of eleven equations 
 for the ten unknown functions  $ {\cal U} = \{\Omega, \Omega^\prime, \Omega^{\prime \prime},\Omega^{\prime \prime \prime}, \Omega^\cdot ,\Omega^{\cdot \cdot}, \Omega^{\cdot \cdot \cdot}, \Omega^{\cdot  \prime}, \Omega^{\cdot \cdot \prime},  \Omega^{\cdot \prime \prime}\}$.
This means that one should be able to obtain a new constraint from it, valid everywhere in the bulk. 
Namely one gets  
\beq  \nonumber
\dot{\tilde{A}}_{yy} + \frac{\dot{a}}{a} \tilde{A}_{yy} + 2 \dot{\tilde{A}} + 2  \frac{\dot{a}}{a}\tilde{A}
+ n^\prime \tilde{A}_y + \frac{a ^\prime n}{2a} \tilde{A}_y + \frac{n}{2}  \tilde{A}_y^\prime =
-\frac{\dot{\Omega}}{2 a} \left( \frac{\Lambda_5}{6} + \frac{n^{\prime \prime}}{n} \right),
\eeq
where the right hand side of the above equation vanishes, when taken into account
the equations of motion for the background leading to (\ref{nppB}). 
If we subtract to this equation two times the time derivative of the constraint (\ref{GQ1}), we obtain the equation 
\beq \label{GQ2}
-\dot{\tilde{A}}_{yy} + \frac{\dot{a}}{a} \tilde{A}_{yy}  + 2  \frac{\dot{a}}{a}\tilde{A}
+ n^\prime \tilde{A}_y + \frac{a ^\prime n}{2a} \tilde{A}_y + \frac{n}{2}  \tilde{A}_y^\prime
-2 \dot{\tilde{\cal R}} = 0,
\eeq
which will be used latter. Removing equation (\ref{EQ5}) from the system  $({\cal S}_0)$, we get 
 a system $({\cal S})$ of ten equations for the ten unknowns ${\cal
   U}$ with second members given by gauge invariant variables and
 their derivatives. This system has a non vanishing determinant in the
 bulk and can be inverted to give $\Omega$ and its derivatives as a
 function of the gauge invariant variables. E.g. one obtains the
 following expression for $\Omega$ (which has been simplified using
 the equations of motion of the background metric)
\beq 
\frac{2 \da \paBDL}{3 a^3} \Delta^2 \Omega &=& \Lc a^2 \da \left( - 2\dot{a}_\bb \tilde{A}_y  +a^\prime  \tilde{A}_{yy} - a \tilde{A}^\prime \right) \nonumber \\
&&+ \dot{a}a^\prime \left( 4 \Delta(\tilde{A} + \tilde{A}_{yy} ) + 6 \dot{a}^2_\bb ( \tilde{A} + 2 \tilde{A}_{yy} ) - 3 \dot{a}_\bb a^\prime \tilde{A}_y\right) \nonumber \\ &&
+ 6 a \dot{a}^2_\bb  \left(2 \ddot{a}_\bb \tilde{A}_y + \dot{a} \tilde{A}^\prime + 2 a^\prime \tilde{A}^\cdot \right) \nonumber \\
&& + 3 a \dot{a}_\bb a^\prime \left(2\ddot{a}_\bb(\tilde{A}-\tilde{A}_{yy}) + 2\dot{a}  \tilde{A}_y^\prime + a^\prime  \tilde{A}_y^\cdot \right),
\label{OmeGIV}
\eeq
which shows in particular that $\Omega$ is  5D-gauge invariant.

In a similar manner to the derivation of equation 
 (\ref{GQ2}), one deduces from the equation
\beq \nonumber
\tilde{\cal R}^\prime=\left\{ \frac{1}{6a} \left( \Omega^{\prime \prime} - \frac{1}{n^2} \ddot{\Omega} + \frac{\Lc}{3} \Omega + \frac{\dot{n}}{n^3} \dot{\Omega} + \frac{n^\prime}{n} \Omega^\prime \right)\right\}^\prime, 
\eeq
the following relation 
\beq \nonumber
\frac{n a^\prime - a n^\prime}{an} \tilde{A} - \tilde{A}^\prime + \frac{2 a^\prime n + a n^\prime}{an} \tilde{A}_{yy} - \frac{\dot{a}}{2 a n} \tilde{A}_y - \frac{1}{2 n} \tilde{A}_y^\cdot - 2 \tilde{\cal R}^\prime = -\frac{\Omega^\prime}{2 a} \left( \frac{\Lambda_5}{6} + \frac{n^{\prime \prime}}{n} \right),
\eeq
which leads to the constraint
\beq \label{GQ3} 
\frac{n a^\prime - a n^\prime}{an} \tilde{A} - \tilde{A}^\prime + \frac{2 a^\prime n + a n^\prime}{an} \tilde{A}_{yy} - \frac{\dot{a}}{2 a n} \tilde{A}_y - \frac{1}{2 n} \tilde{A}_y^\cdot - 2 \tilde{\cal R}^\prime = 0.
\eeq

Taking the limiting values of the constraints (\ref{GQ1}), (\ref{GQ2}) and (\ref{GQ3}) 
 on the brane, and replacing in it the gauge independent variable by their expression as a function of effective matter and Weyl's fluid given by equations (\ref{Atb}-\ref{Rtbp}) one obtains the following three equations
\label{PCGI14p25} 
\beq
\label{EQM1s}
 \left\{\Phi - \Psi + a^2 \kqd \delta \pi_\EE -
 \frac{\kcq}{12 } a^2 \left( 3P + \rho\right) \delta \pi \right\}_\bb &=& 0  \\
 \left\{ 2 \dot{\Psi} + 2 \frac{\dot{a}}{a} \Phi
+ \frac{\kcq}{6} \delta q \rho  + \kqd \delta q_\EE \right\}_\bb \label{EQM4s} &=&0 \\
\left\{ \delta \dot{q}+   3 \frac{\dot{a}}{a}  \delta q  + \delta P +\left( P + \rho \right) \Phi + \frac{2}{3}  \Delta \delta \pi  \right\}_\bb &=&0 
\label{EQM3s}.
\eeq
Those three equations are respectively the trace free part of the 
$ij$ component of the 
perturbed effective Einstein equation (\ref{Perteffein}), the $0i$ component of the same  equation, and the 
conservation equation for the effective matter momentum $\delta q$ (that is to say the $i$ component of the perturbed equation (\ref{consS})). 
Those equations are derived  in appendix \ref{appD} from equations (\ref{effein}) and (\ref{consS}).
We mention here that equations (\ref{EQM1s}) and (\ref{EQM4s}) can be used to obtain an expression of 
 $\delta \pi_\EE$ and $\delta q_\EE$ as a function of effective matter perturbations. The latter can then be used in equations (\ref{Atb}-\ref{Rtbp}) to express the left-hand side of those equations solely in terms of the effective matter and induced metric perturbations.

We would like now to derive the remaining perturbed effective Einstein's and conservation equations. 
Those imply constraints involving second derivatives of the gauge invariant variables. 
To be more precise, we will need the limiting values on the brane of $\tilde{A}^{\cdot \prime}, \tilde{A}_y^{\cdot \cdot}, 
\tilde{A}_y^{\cdot \prime}$ as well as $\tilde{R}^{\prime \prime}, \tilde{A}^{\prime \prime}$. The first three limiting values are simply given by time derivatives of equations (\ref{Atb}), (\ref{Atyb}) and  (\ref{Atybp}). However, 
computing the limiting values of $\tilde{R}^{\prime \prime}, \tilde{A}^{\prime \prime}$ on the brane require
knowing $E^{\prime \prime \prime}$ on the brane, as can be seen from the expressions (\ref{A}) and (\ref{R}).
This can be obtained using the first constraint (\ref{GQ1}), leading to\label{[PCGI ANN 6 P 6]}
\beq
\tilde{A}_{yy \bb}^\prime &=& - \tilde{A}^\prime_\bb - \tilde{R}^\prime_\bb \nonumber \\ \nonumber
&=& \left\{ a^2 E^{\prime \prime \prime} + 4 a^\prime a E^{\prime \prime} + (2a a^\prime)^\prime E^\prime \right\}_\bb
\eeq
where the last equality comes from (\ref{Ayy}). 
Solving for $E^{\prime \prime \prime}_\bb$ one obtains (using again the junctions conditions of section \ref{JUNCON}  as well as the background equation of motion of appendix \ref{appA}) \label{[PCGI ANN 12]}
\beq
a^2_\bb E^{\prime \prime \prime}_\bb &=& \left\{ -\frac{\kcd}{2} \delta P - \frac{\kcd}{6} \delta \rho  - \kcd \frac{\dot{a}}{a} \delta q - \kcd \dot{\delta q}  \nonumber \right.\\ 
&&-\frac{\kcd}{2} a^2 \ddot{\delta \pi} - \frac{3\kcd}{2} a \dot{a} \kcd \dot{\delta \pi} \nonumber \\
&& - \frac{\kcd}{72} \left\{ 24 a^2 \Lc + \kcq a^2 \left( 11 \rho^2 + 12 P \rho + 9 P^2 \right) + 12 \Delta \right\} \delta \pi \nonumber \\
&& \left.+ \frac{\kcd}{2} \kqd a^2 \left( \rho -P \right) \delta \pi_\EE \right\}_\bb \label{Ebppp},
\eeq
which allows to compute ${R}^{\prime \prime}_\bb$ and $\tilde{A}^{\prime \prime}_\bb$. 
The next step of our derivation is to extract 
the fourth derivatives $\Omega^{\cdot \cdot \cdot \cdot}, \Omega^{\cdot \cdot \cdot \prime}, \Omega^{\cdot \cdot \prime \prime}, \Omega^{\cdot \prime \prime \prime}, \Omega^{\prime \prime \prime \prime}$ 
from the expression of $\tilde{A}_{y}^{\cdot \cdot}$ and  $\tilde{A}_{y}^{\cdot \prime}$
as a functions of $\Omega$ and its derivatives
(that one gets from equation (\ref{EQ6}))
, and from the second derivatives $\partial_t \partial_t$, $\partial_y \partial_t$, $\partial_y \partial_y$ of the master equation (\ref{FQ1}). Having done so one can insert those fourth derivatives in the expressions of $\tilde{A}^{\cdot \prime}$, $\tilde{R}^{\prime \prime}$ and $\tilde{A}^{\prime \prime}$ as a function of $\Omega$ and its derivatives. This lead to an expression for each of the quantities 
$\tilde{A}^{\cdot \prime}$, $\tilde{R}^{\prime \prime}$ and $\tilde{A}^{\prime \prime}$ which does only contain derivatives of $\Omega$ of order lower or equal to three. One can then use, as in the derivation of constraints (\ref{GQ1}), (\ref{GQ2}) and (\ref{GQ3}) the system $({\cal S})$  to obtain an expression in the bulk relating each of the functions
$\tilde{A}^{\cdot \prime}$, $\tilde{R}^{\prime \prime}$ and $\tilde{A}^{\prime \prime}$ to  $\tilde{A}_{y}^{\cdot \cdot}, \tilde{A}_{y}^{\cdot \prime}, \tilde{A}^{\cdot}, \tilde{A}^\prime, \tilde{A}_y^\cdot, \tilde{A}_y^\prime, \tilde{A}_{yy}^\cdot, \tilde{A}, \tilde{A}_y, \tilde{A}_{yy}$ as well as to the background metric. 
The last step is to take the limiting values of those expressions on the brane, and to replace in it 
all the gauge independent variables, as well as their derivatives, by their expressions as a function of brane effective matter and Weyl's fluid. 
The expression for $\tilde{A}^{\cdot \prime}$ leads to 
\beq \label{EQM5s} \left\{
\delta \dot{\rho} + 3 \frac{\dot{a}}{a} ( \delta \rho + \delta P) + \frac{\Delta}{a^2} \delta q - 3 (P+\rho) \dot{\Psi} \right\}_\bb = 0,
\eeq
which is the $0$ component of the conservation equation (\ref{consS}), or in other words the effective matter energy density conservation equation.
The expression for $\tilde{R}^{\prime \prime}$ leads to 
\beq \label{EQM6s} \left\{
6 \frac{\dot{a}}{a} \left( \dot{\Psi} + \frac{\dot{a}}{a} \Phi \right) - \frac{2}{a^2} \Delta \Psi\right\}_\bb = - \frac{\kcq}{6} \rho \delta \rho - \kqd \delta \rho_\EE,
\eeq 
which is the $00$ component of the perturbed effective Einstein's equation (\ref{Perteffein}). While the computing on the brane 
$\tilde{A}^{\prime \prime}+ 2 \tilde{R}^{\prime \prime}$ leads to 
\beq 
\frac{\kcq}{6}  \left((\rho+P)  \delta \rho + \rho \delta P \right) + \kqd \delta P_\EE  &=& \nonumber  \left\{2\left(2  \frac{\ddot{a}}{a} + \frac{\dot{a}^2}{a^2} \right)\Phi + 2 \frac{\dot{a}}{a} \dot{\Phi} +  
2 \ddot{\Psi} \right. \\
&& \left. + 6 \frac{\dot{a}}{a} \dot{\Psi} + \frac{2}{3 a^2} \Delta \left( \Phi - \Psi\right) \right\}_\bb, \label{EQM8s} 
\eeq
which is the trace of the $ij$ components of effective Einstein's equation
(\ref{Perteffein}). 
The last two equations are the  
equations coming from perturbing the Weyl's fluid conservation equation (\ref{consweyl}). They  
can then be derived from equations (\ref{EQM1s}), (\ref{EQM4s}), (\ref{EQM3s}), (\ref{EQM5s}), (\ref{EQM6s}), (\ref{EQM8s}),
(as a  consequence of Bianchi identities) and read  
 \beq \left\{
\dot{\delta \rho_\EE} + 3 \frac{\dot{a}}{a} \left(\delta \rho_\EE + \delta P_\EE \right)  + \frac{\Delta}{a^2} \delta q_\EE \right\}_\bb
 &=&  0, \label{EQM7s} 
\eeq
and 
\beq \left\{ 
 \dot{\delta q_\EE} + 3 \frac{\dot{a}}{a} \delta q_\EE
+  \delta P_\EE + \frac{2}{3} \Delta \delta \pi_\EE \right\}_\bb=  \left\{\label{EQM9s}  
\frac{\kcq}{6 \kqd}(P+\rho)  \left( 3 \frac{\dot{a}}{a} \delta q + \Delta \delta \pi  -\delta \rho  \right) \right\}_\bb.
\eeq
We have thus showed at that point that all the scalar components of the perturbed effective Einstein's 
equations (\ref{effein}) and conservation equations (\ref{consS}) and (\ref{consweyl}) can be derived from the Master equations and the junction conditions. 
We now discuss the form of the boundary condition for $\Omega$ on the brane. 

\section{Boundary condition on the brane from real matter equation of state} \label{BOUND}
We first note that inserting the expressions  (\ref{Atb}-\ref{Rtbp}) (as well as using equation (\ref{EQM3s})) in the limiting value on the brane of equation 
(\ref{OmeGIV}), one gets the following expression  for $\Delta^2 \Omega_\bb$
\beq
\Delta^2 \Omega_\bb  &=& \left\{- 18 a^4 \dot{a} \left( \dot{\Psi} + \frac{\dot{a}}{a} \Phi \right) + 6 a^3 \Delta \Psi - \frac{\kcq}{2} a^5 \rho \delta \rho\right\}_\bb \label{Omat}.
\eeq
Using similarly the expression for $\Omega^\prime$ obtained from $({\cal S})$ we get 
\beq
\Delta \Omega^\prime_\bb &=& \left\{ -\frac{\kcd}{6} \rho \Delta \Omega + \kcd a^3 \delta \rho - 3 \kcd a^2 \dot{a} \delta q - \kcd a^3 \Delta \delta \pi \right\}_\bb \label{Opmat}.
\eeq
Similar equations were obtained by Mukohyama \cite{Mukoh2}  and used by him  
get an integro-differential equation for the effective matter and induced metric perturbation on the brane
One may be tempted to consider them as a way to fix a boundary condition for the differential operator ${\cal D}_\Delta$. This is however not the case since the right hand side of those equations is not known as a function of time. 
Moreover, these equations do not allow to
easily disentangle the extra physical conditions that  have to be imposed 
to get a boundary condition, nor to discuss the well-poseness issue. To obtain the latter we wish to express the effective matter and induced metric perturbations as functions of the master variable, as we now do, discussing first the case of Weyl's fluid perturbations.

\subsection{Relation between  Weyl's fluid, effective matter perturbations and the master variable}
\subsubsection{Weyl's fluid}

From the equations (\ref{EQ4}) and (\ref{Atyyb}) one easily finds
an expression for the Weyl's fluid anisotropic stress $\delta \pi _\EE$ given by  \label{[PCGI ANN 14]}
\beq
\delta \pi_\EE = \frac{1}{6 \kqd a_\bb^3}\left\{3 \ddot{\Omega} - 3 \frac{\dot{a}}{a} \dot{\Omega} - \frac{\Delta}{a^2} \Omega - \frac{3}{2} \kcd (P+\rho) \Omega^\prime \right\}_\bb \label{DPIEO}.
\eeq
The energy density perturbation of the Weyl's fluid can similarly be derived from equations (\ref{Omat}) and (\ref{EQM6s}), one gets 
\label{[PCGI 23 p25]}
\beq \label{DREO}
\delta \rho_\EE = \left\{\frac{\Delta^2 \Omega}{3 \kqd a^5} \right\}_\bb.
\eeq
The perturbation of the Weyl's fluid pressure is then given by equation (\ref{Weyleq}), whereas $\delta q_\EE$ can be obtained
using the conservation equation (\ref{EQM7s}) and reads
\beq \label{DQEO}
\delta q_\EE = \left\{ \frac{1}{3 \kqd a^3} \left(\frac{\dot{a}}{a}  \Delta  \Omega - \Delta \dot{\Omega} \right) \right\}_\bb.
\eeq
Those expressions can be used to compute the perturbations of the Weyl's fluid energy momentum tensor 
once $\Omega$ is known as function of time, and e.g. compute the Sachs-Wolfe effect \cite{LMSW}\footnote{The Sachs-Wolfe effect can also be computed 
directly, under the same hypothesis, using the expressions for $\Phi_\bb$ and $\Psi_\bb$ given in (\ref{PHIO}), (\ref{PHI1}) and (\ref{PSIO}), (\ref{PSI1}).}. They can also be compared with similar expressions 
that will be obtained below for effective (or real) matter and induced metric perturbations in order to test various approximations done in the literature to deal with the Weyl's fluid (like in \cite{Leong:2001qm}).

\subsubsection{Effective matter}
In analogy to the above expression for the Weyl's fluid perturbations,  we wish to obtain expressions for the effective matter energy momentum perturbations and induced metric on the brane as  functions of the master variable $\Omega$ and its derivatives. 
To achieve this, one can think to use 
 the system (${\cal S}$), which  can also be  considered as a system $({\cal S}^\prime)$ in the 10 unknown functions $\delta \rho, \delta P, \delta q, \delta \pi, \delta \dot{\pi}, \delta \ddot{\pi}, 
\Psi, \dot{\Psi}, \Phi, \dot{\Phi}$  once one replaces the gauge invariant variables (and their relevant derivatives) by their limiting values on the brane as functions of the effective matter and induced metric perturbation. 
This system $({\cal S}^\prime)$  is however not invertible, but  one can obtain from it the following expressions
\beq \nonumber 
\Phi_\bb &=& \frac{1}{6 a_\bb} \left\{ \left( \frac{2 \Delta}{a^2} - \frac{\Lc}{2} \right) \Omega  + 6 \frac{\dot{a}}{a} \dot{\Omega} - 3 \ddot{\Omega} + \frac{\kcd}{2} \left( 3P + 4 \rho \right) \Omega^\prime \right. \\ \label{phbPHI}
&&\left. + \frac{\kcq}{2}  a^3 \left( 3P + 2 \rho \right) \delta \pi \right\}_\bb, \\  \label{psbPHI}
\Psi_\bb &=& \frac{1}{6 a_\bb} \left\{ \left( \frac{\Delta}{a^2} - \frac{\Lc}{2} \right) \Omega  + 3 \frac{\dot{a}}{a} \dot{\Omega}  + \frac{\kcd}{2}  \rho  \Omega^\prime        + \frac{\kcq}{2}  a^3 \rho \delta \pi     \right\}_\bb, \\ \nonumber 
\delta \rho &=& \frac{1}{6 a_\bb} \left\{ \rho \frac{\Delta \Omega}{a^2} + 3 \frac{\dot{a}}{a} \left( 3P + 2 \rho\right)  \dot{\Omega} + 6 \frac{\Delta \Omega^\prime}{\kcd a^2}  \right.\\ && \left.  - 18 \frac{\dot{a} \Omega^{\cdot \prime} }{ \kcd a} 
+ 6a \Delta \delta \pi - 18 \dot{a} a \left( a \delta \pi \right)^\cdot  \right\}_\bb \label{drPHI}, \\ \label{dqPHI}
\delta q &=& \frac{1}{6a_\bb} \left\{ \left(3P + 2 \rho \right) \dot{\Omega} - \frac{6}{\kcd} \Omega^{\cdot \prime} - 6 a^2 \left(a \delta \pi \right)^\cdot \right\}_\bb,  \\ \nonumber 
\delta P &=&  \frac{1}{6a_\bb} \left\{ \left( P+ \rho \right) \left(- \frac{2 \Delta}{a^2} + \frac{\Lc}{2} \right) \Omega  
- \left( 3 \dot{P} + \frac{\dot{a}}{a} \left( 4 \rho + 6 P \right) \right) \dot{\Omega} \nonumber \right. \\
&& \left. + \rho \ddot{\Omega}   -\frac{\kcd}{2} \left(P+ \rho \right) \left( 3P + 4 \rho \right) \Omega^\prime  + \frac{12 }{\kcd } 
\frac{\dot{a}}{a}
\Omega^{\cdot \prime} + \frac{6}{\kcd} \Omega^{\cdot \cdot \prime} \right. \nonumber \\
&& \left.    - a^3 \left[
4 \frac{\Delta}{a^2} +  4 \Lc   + \frac{\kcq}{6} \left( 9 P^2 + 9 P \rho + 4 \rho^2 \right) \right] \delta \pi + 6 (a^3 \delta \pi )^{\cdot \cdot} \right\}_\bb .\label{dPPHI}
\eeq
One can verify by inserting those expressions in equations (\ref{EQM1s}), (\ref{EQM4s}), (\ref{EQM3s}), (\ref{EQM5s}), (\ref{EQM6s}), (\ref{EQM8s}), 
(\ref{EQM7s}) and (\ref{EQM9s}) that those equations,  namely the  perturbed effective Einstein's equations and conservation equations,  are identically satisfied whatever $\Omega$. This means that  the latter do not  contain more  informations than the above equations. 
The fact that $({\cal S}^\prime)$ is not invertible  could have been expected counting the number of 
extra equations  one needs to get a closed system (or  a boundary condition)  as we now explain.

 Let us first recall what is happening for usual 4D cosmological
 perturbations (the linearized 4D Einstein's equations and
 conservation equations are reminded in appendix \ref{4Dpert}).
 In this case if one considers only one matter 
fluid as a source in the Einstein's equations, and using a decomposition of its energy momentum tensor of the form given in equations (\ref{dTM00}-\ref{dTMij}), one gets a system of equations from the perturbed Einstein's equations which does not close. To close the system and solve for the evolution of cosmological perturbations once initial conditions are provided, one needs to add more equations, e.g.  equations of state for the matter fluid. The simplest case considered in general is the case of a perfect fluid, which 
implies
\beq \label{perfect}
\delta \pi_\MM = 0,
\eeq  
with adiabatic perturbations obeying  
\beq \label{adiab}
\delta P_\MM = c_s^2 \delta \rho_\MM,
\eeq 
where $c_s^2$ is the sound velocity defined by 
\beq \label{sound}
c_s^2 = \frac{\dot{P}_\MM}{\dot{\rho}_\MM}.
\eeq
The two conditions (\ref{perfect}) and (\ref{adiab}) are sufficient to
close the system of perturbed Einstein equations,  leading to the
evolution equation for the gravitational potential $\Phi$ 
(defined as in equation (\ref{4Dlinepert}))
\label{PCGI 24 page 10}
\beq  \label{EVOL}
\ddot{\Phi}  + (4 + 3 c_s^2) \frac{\dot{a}}{a} \dot{\Phi} + \left[ 2 \frac{\ddot{a}}{a} + \frac{\dot{a}^2}{a} (1 + 3 c_s^2) \right] \Phi - \frac{c_s^2}{a^2} \Delta \Phi = 0,
\eeq 
and all the other relevant quantities can be deduced from the knowledge of $\Phi$ as a function of time.

Turning back to the brane perturbations, if one would be able to get an expression for $\delta \pi_\MM, \delta \rho_\MM$ and $\delta P_\MM$ as a function of $\Omega$ and its derivatives on the brane (that is to say if $({\cal S'})$ would have been invertible), then imposing the two conditions (\ref{perfect}) and (\ref{adiab}) would lead to two independent boundary conditions on the brane, which is too much, once initial conditions are supplied on a 
suitable hyper-surface in the bulk. From the above discussion one can however expect as well to have to use two extra equations  
(e.g. equations of state on the matter fluid on the brane) in order to get a solvable problem for the evolution of cosmological perturbations on the brane. The first of those extra equations can be though of as enabling to express the matter fluid energy momentum perturbations as a function of the master variable and its derivative (namely to invert $({\cal S'})$), while the other leads to a single boundary condition on the brane.  This will be done explicitly below for the two models described in section \ref{MOGEO}. 

\subsection{Boundary condition for {\it real} perfect fluid with adiabatic perturbations} \label{bound}
\label{PCGI Ann 15}
In the case of the RS model (model (I) in section \ref{MOGEO}), one has simply the following relations between effective and real matter energy momentum perturbations
\beq \nonumber
\delta \rho &=& \delta \rho_\MM, \\ \nonumber
\delta P &=& \delta P_\MM, \\ \nonumber
\delta q &=& \delta q_\MM ,\\ \nonumber
\delta \pi &=& \delta \pi_\MM,  
\eeq
as can be seen from equation (\ref{SRS}). 
We now specialize to the case where the real matter is a perfect fluid. In this case the anisotropic stress $\delta \pi_\MM$ vanishes 
and one gets from equations (\ref{phbPHI}-\ref{dPPHI}) 
\beq
\Phi_\bb &=& \frac{1}{6 a_\bb} \left\{ \left( \frac{2 \Delta}{a^2} - \frac{\Lc}{2} \right) \Omega  + 6 \frac{\dot{a}}{a} \dot{\Omega} - 3 \ddot{\Omega} + \frac{\kcd}{2} \left( 3P + 4 \rho \right) \Omega^\prime \right\}_\bb \label{PHIO}, \\
\Psi_\bb &=& \frac{1}{6 a_\bb} \left\{ \left( \frac{\Delta}{a^2} - \frac{\Lc}{2} \right) \Omega  + 3 \frac{\dot{a}}{a} \dot{\Omega}  + \frac{\kcd}{2}  \rho  \Omega^\prime \right\}_\bb \label{PSIO}, \\
\delta \rho_\MM &=& \frac{1}{6 a_\bb} \left\{ \rho \frac{\Delta \Omega}{a^2} + 3 \frac{\dot{a}}{a} \left( 3P + 2 \rho\right)  \dot{\Omega} + 6 \frac{\Delta \Omega^\prime}{\kcd a^2}  - 18 \frac{\dot{a} \Omega^{\cdot \prime} }{ \kcd a}  \right\}_\bb \label{DRO}, \\
\delta q_\MM &=& \frac{1}{6a_\bb} \left\{ \left(3P + 2 \rho \right) \dot{\Omega} - \frac{6}{\kcd} \Omega^{\cdot \prime} \right\}_\bb \label{DQO} ,\\ 
\delta P_\MM &=&  \frac{1}{6a_\bb} \left\{ \left( P+ \rho \right) \left(- \frac{2 \Delta}{a^2} + \frac{\Lc}{2} \right) \Omega  
- \left( 3 \dot{P} + \frac{\dot{a}}{a} \left( 4 \rho + 6 P \right) \right) \dot{\Omega} \nonumber \right. \\
&& \left. + \rho \ddot{\Omega}   -\frac{\kcd}{2} \left(P+ \rho \right) \left( 3P + 4 \rho \right) \Omega^\prime  + \frac{12 }{\kcd } 
\frac{\dot{a}}{a}
\Omega^{\cdot \prime} + \frac{6}{\kcd} \Omega^{\cdot \cdot \prime} \right\}_\bb \label{DPO}.
\eeq
These expressions allow to get a boundary condition for the operator ${\cal D}_\Delta$ on the brane, using an extra 
physical condition on the real matter and induced metric perturbations. In the case of adiabatic perturbations, the latter is given by equation (\ref{adiab}). 
If we insert in (\ref{adiab}) the expressions for $\delta \rho_\MM$ and $\delta P_\MM$ given by 
equations (\ref{DRO}) and (\ref{DPO}) we get a  boundary condition
for ${\cal D}_\Delta$ which is  expressible by 
\beq \label{boundRS}
F_{RS}(\Omega)_\bb+ G_{RS}(\Omega^\prime)_\bb =0, 
\eeq
where $F_{RS}$ and $G_{RS}$ are polynomials of the cosmic time-derivative, $\partial_t$, with cosmic time dependent coefficients which are known
 from equations (\ref{sound}), (\ref{DRO}) and (\ref{DPO}) as well as from the background solution.  

We now turn to the DGP model. In this case, one gets from  equation (\ref{SDGP})  the following expressions
for the effective matter perturbations as functions of real matter and induced metric perturbations
\label{PCGI,Ann18-19}
\beq \nonumber
\delta \rho &=& \delta \rho_\MM + \frac{1}{\kqd}\left\{ 6 \frac{\dot{a}}{a}\left(\dot{ \Psi} + \frac{\dot{a}}{a} \Phi \right) - 2\frac{\Delta}{a^2} \Psi \right\}_\bb, \\ \nonumber
\delta q &=& \delta q_\MM + \frac{2}{\kqd} \left\{ \frac{\dot{a}}{a} \Phi + \dot{\Psi} \right\}_\bb, \\ \nonumber
\delta P &=& \delta P_\MM - \frac{1}{\kqd} \left\{  \left( 4 \frac{\ddot{a}}{a} +2 \frac{\dot{a}^2}{a^2} \right) \Phi + 2 \frac{\dot{a}}{a} \dot{\Phi} + 2 \ddot{\Psi} + 6 \frac{\dot{a}}{a} \dot{\Psi} + \frac{2 \Delta }{3 a^2} (\Phi - \Psi) \right\}_\bb , \\ \nonumber
\delta \pi &=& \delta \pi_\MM + \left\{ \frac{\Phi - \Psi} {\kqd a^2} \right\}_\bb.
\eeq 
Substituting the above expressions for the effective matter perturbations $\delta \rho,\delta q, \delta P$ and
 $\delta \pi$ in equations  (\ref{phbPHI}-\ref{dPPHI}),
and replacing $\delta \pi_\MM$ by zero, one 
 obtains for a perfect fluid on a DGP brane
\beq
\Phi_\bb &=& \frac{1}{6 a_\bb} \left\{ \frac{\Delta \Omega}{a^2}  \CC^{\Phi}_{\Delta (0,0)}
+ H \dot{\Omega}  \CC^{\Phi}_{(1,0)} +\ddot{\Omega}   \CC^{\Phi}_{(2,0)} + 
 H \Omega^\prime \CC^{\Phi}_{(0,1)} \right\}_\bb  \label{PHI1},\\
\Psi_\bb &=& \frac{1}{6 a_\bb} \left\{ \frac{\Delta \Omega}{a^2}  \CC^{\Psi}_{\Delta (0,0)}
+ H \dot{\Omega}  \CC^{\Psi}_{(1,0)} +  \ddot{\Omega}  \CC^{\Psi}_{(2,0)}+ 
H \Omega^\prime  \CC^{\Psi}_{(0,1)}  \right\}_\bb \label{PSI1},\\
\kqd \delta \rho_\MM& =& \frac{H^2}{2 a_\bb} \left\{
 \frac{\Delta \Omega}{a^2}  \CC^\rho_{\Delta (0,0)} +  \frac{\Delta^2 \Omega}{H^2 a^4}  \CC^\rho_{\Delta^2 (0,0)}
+  H \dot{\Omega} \CC^\rho_{(1,0)} +   \ddot{\Omega}  \CC^\rho_{(2,0)} 
+  \frac{\Delta \ddot{\Omega}}{H^2 a^2}  \CC^\rho_{\Delta (2,0)}  \nonumber \right.\\
&& \left.+ \frac{\Omega^{\cdot \cdot \cdot}}{H} \CC^\rho_ {(3,0)} + 
H \Omega^\prime \CC^\rho_{(0,1)} +  \frac{\Delta \Omega^\prime}{H a^2}  \CC^\rho_{\Delta (0,1)} +  \dot{\Omega^\prime}   \CC^\rho_ {(1,1)}
\right\}_\bb \label{DR1},\\
\kqd \delta q_\MM& =& \frac{H}{6 a_\bb} \left\{
 \frac{\Delta \Omega}{a^2}  \CC^q_{\Delta (0,0)} +
 H \dot{\Omega}  \CC^{q}_{(1,0)} + \frac{\Delta \dot{\Omega}}{H a^2} \CC^{q}_{\Delta (1,0)}
+ \ddot{\Omega}   \CC^{q}_{(2,0)} +  \frac{\Omega^{\cdot \cdot \cdot}}{H} \CC^q_ {(3,0)} \nonumber \right.\\
&&  \left.
+  H \Omega^\prime \CC^q_{(0,1)} +  \dot{\Omega^\prime}   \CC^q_ {(1,1)} \right\}_\bb, \label{DQ1}\\
\kqd \delta P_\MM& =& \frac{H^2}{6 a_\bb} \left\{
 \frac{\Delta \Omega}{a^2}  \CC^P_{\Delta (0,0)} + H \dot{\Omega}  \CC^{P}_{(1,0)} + \frac{\Delta \dot{\Omega}}{H a^2} \CC^{P}_{\Delta (1,0)}
+  \ddot{\Omega}  \CC^P_{(2,0)}
+  \frac{\Delta \ddot{\Omega}}{H^2 a^2}  \CC^P_{\Delta (2,0)}  \nonumber \right.\\
&&  \left. +  \frac{\Omega^{\cdot \cdot \cdot}}{H} \CC^P_ {(3,0)}  + \frac{\Omega^{\cdot \cdot \cdot \cdot}}{H^2} \CC^P_ {(4,0)}
+  H \Omega^\prime \CC^P_{(0,1)} +  \dot{\Omega^\prime}   \CC^P_ {(1,1)} +  \frac{\Omega^{\cdot \cdot \prime}}{H}   \CC^P_ {(2,1)}
\right\}_\bb, \label{DP1}
\eeq
where the coefficients $\CC$ are given in appendix  \ref{expressC}, and our normalization is chosen so that they are dimensionless.
As for the RS model, we obtain from those equations a boundary condition for ${\cal D}_\Delta$ expressible 
as
\beq \label{boundDGP}
F_{DGP}(\Omega)_\bb+ G_{DGP}(\Omega^\prime)_\bb =0, 
\eeq
where $F_{DGP}$ and $G_{DGP}$ are polynomials of the cosmic time-derivative,  $\partial_t$, with time dependent coefficients known from the background solution.  

The boundary conditions (\ref{boundRS}) and (\ref{boundDGP}) 
have an unconventional form, since they involve derivatives of
$\Omega$ and $\Omega^\prime$ along the brane. However one can easily see, in 
the simple case where the master equation in the bulk is a wave
equation, that such boundary conditions lead in this case to a well
posed problem. This is shown explicitely in appendix \ref{appDD}.  We
expect this to be true as well here, since those boundary conditions
and Mukohyama's master equation encodes a priory all the physics of
the problem. This can indeed be verified numerically. 
Equations (\ref{FQ1}) and (\ref{boundRS}) (respectively
(\ref{boundDGP})) should be then 
all what is needed to solve for the evolution of RS (respectively DGP)
brane world cosmological perturbations once initial conditions are
supplied in the bulk as explained in subsection \ref{MASTER} (see also
appendix \ref{appDD}). Those two equations play for the brane world cosmological perturbations an equivalent role to the one played by equation (\ref{EVOL}) for 4D adiabatic cosmological perturbations
of a perfect fluid. Once $\Omega$ is known in the vicinity of the brane, one can compute all the induced metric and real matter perturbations from equations (\ref{PHIO}-\ref{DPO}) (respectively 
(\ref{PHI1}-\ref{DP1})).  We stress again here that one can verify that the real matter, induced metric perturbations defined from those equations in terms of the master variable, as well as the Weyl's fluid perturbations given by equations (\ref{DPIEO}), (\ref{DREO}) and (\ref{DQEO}) verify together identically 
(i.e. whatever the function $\Omega$) the perturbed effective
Einstein's equations (\ref{effein}) and conservation equations
(\ref{consmat}) and (\ref{consweyl}) (which components are gathered in
appendix \ref{B2}). As a last remark, we would like to comment on the GN coordinate system we used to derive the boundary condition. In general, there is no 
warranty that this coordinate system will not break down at some distance in the bulk. However, as we have seen above (from equation (\ref{OmeGIV}))
 $\Omega$ is a gauge invariant quantity. Since the boundary condition  (\ref{boundRS}) (or (\ref{boundDGP})) is here expressed solely in terms of the master variable, it keeps the same form in a non GN coordinate system and can always be imposed on the background brane trajectory. 

\section*{Acknowledgments}
We thank G.~Dvali, D.~Langlois, A.~Lue, K.~Malik, M.~Porrati, R.~Scoccimarro,
J.~Shata and M.~Zaldarriaga for useful discussions. 
This work is sponsored by NSF Award PHY 9803174 and by David and Lucile Packard Foundation Fellowship 99-1462.  



\appendix

\section{Some results for the background  metric} \label{appA0}
In this appendix we give some results for the background bulk and brane metric. 
\subsection{Bulk relations}
\label{PCGI 24 p15}
The Einstein's equations (\ref{einstein}) together with the cosmological ansatz (\ref{backmet}) lead to the following relations valid in the bulk \cite{Binetruy:2000ut} (for a spatially flat metric on the brane)
\beq
G_{00}^{(5)} &=& 3\left\{ \frac{\dot{a}^2}{a^2} - n^2 
\left(\fppa + \frac{a^{\prime 2}}{a^2} \right)  \right\},  \nonumber 
 \\
&=& n^2 \Lc \label{ein00} ,\\
G_{ij}^{(5)} &=& \nonumber 
{a^2} \eta_{ij}\left\{\fpa
\left(\fpa+2\fpn\right)
+2\fppa+\fppn 
+\frac{1}{n^2}  \left( \fda \left(-\fda+2\fdn\right)-2\fdda
 \right) \right\},\\
&=& - a^2 \Lc \eta_{ij} ,
\label{einij} \\ \nonumber 
G^{(5)}_{05} &=&  3\left(\fpn \fda  - \frac{\dot{a}^{\prime}}{a}
 \right), \\
&=& 0 ,
\label{ein05} \\ \nonumber 
G^{(5)}_{55} &=& 3\left\{ \fpa \left(\fpa+\fpn \right) - \frac{1}{n^2} 
\left(\fda \left(\fda-\fdn \right) + \fdda\right) \right\}, \\
&=& - \Lc .
\label{ein55}
\eeq
The explicit form of the metric in the GN coordinate system
(\ref{backmet}) can be obtained from those equations
\cite{Binetruy:2000hy}. We will not reproduce here the general
results, but  give here first the metric for the case of a Minkowskian bulk (with vanishing bulk cosmological constant and Weyl's tensor). In this case one finds (for a brane with positive effective tension)
\beq
a &=& a_\bb - |y| \dot{a}_\bb \nonumber, \\  
n&=& 1 - |y| \ddot{a}_\bb / \dot{a}_\bb \label{bulkmetmink}.
\eeq
In the above equation the unknown function of cosmic time $a_\bb$
has to be determined solving the brane Friedmann's equations which are
given in subsection \ref{brarel}. In the more general case where one
only assumes a vanishing bulk Weyl's tensor, 
one gets the following relations valid everywhere in the bulk
\beq
\frac{n^{\prime \prime}}{n} &=& - \frac{\Lc}{6} \label{nppB}, \\
\frac{a^{\prime \prime}}{a} &=& - \frac{\Lc}{6} \label{appB} .
\eeq

\subsection{Brane relations} \label{brarel}
Using the Junction conditions (\ref{backjun}) and (\ref{junn}) and solving for the bulk metric one obtains the following brane Friedmann's equations \cite{Binetruy:2000hy}
\label{PCGI Ann 11}
\beq \nonumber
\frac{\dot{a}^2_\bb}{a^2_\bb} &=& \frac{\Lc}{6} + \frac{\kcq}{36} \rho^2 ,\\ \nonumber
\frac{\ddot{a}_\bb}{a_\bb} &=&  \frac{\Lc}{6} -  \frac{\kcq}{36} \rho(2\rho+3P),
\eeq
together with the conservation equation (\ref{consback}).
The relations between effective matter energy density $\rho$ and
pressure $P$, entering into those equations, and
real matter energy density $\rho_\MM$ and pressure $P_\MM$ 
  depend on the model considered. In the case of the RS model (model(I)) one gets from equation (\ref{SRS}) 
\beq
\rho &=& \rho_\MM  + \Lq, \label{RRS}\\
P &=& P_\MM - \Lq. \label{PRS}
\eeq
For the DGP model (model (II)) one has \cite{Deffayet:2001uy}
\label{PCGI24/p18}
\beq
\rho &=& \rho_\MM - \frac{3}{\kqd} \frac{\dot{a}^2}{a^2}, \label{RDGP}\\ 
P &=& P_\MM + \frac{1}{\kqd} \left( \frac{\dot{a}^2}{a^2} + 2 \frac{\ddot{a}}{a} \right). \label{PDGP}
\eeq 
Using the above formulas, one can then solve for the explicit time dependence of the metric (see  \cite{Binetruy:2000hy} for the RS model, and \cite{Deffayet:2001uy,Deffayet:2001aw} for the DGP model).

\label{appA}

\section{Scalar cosmological perturbations in 4D GR  and on the brane}
In this appendix we first gather some results on standard 4D scalar cosmological perturbations
 and then give the analogous equations 
for brane worlds, which can be derived linearizing  the effective Einstein's equations on the brane (\ref{effein})
and conservation equations (\ref{consS}) and (\ref{consweyl}) (see e.g. \cite{Bridgman:2001mc}).

\subsection{4D cosmological perturbations}
\label{4Dpert}
We  gather here some useful results concerning standard 4D scalar cosmological perturbations, we refer to the numerous review articles on the subject for more details (see e.g. \cite{Mukhanov:1992me}). 
We work in the longitudinal gauge, where the (4D) perturbed line element reads
\beq \label{4Dlinepert}
ds^2 = -(1+2 \Phi) dt^2 + a^2(t)(1-2 \Psi) \delta_{ij} dx^i dx^j, 
\eeq
with $a(t)$ being the background scale factor. 
The matter energy momentum tensor   $T^\MM_{\mu \nu}$ is given in the background by 
\beq \nonumber
T^{\MM \mu}_{ \nu} = \mbox{ diag} ( -\rho_\MM, P_\MM,P_\MM,P_\MM), 
\eeq
while the perturbed matter energy momentum tensor is defined as in equations (\ref{dTM00}-\ref{dTMij}). A  
scalar gauge transformation given by 
\beq
t &\rightarrow& t+\delta t  \label{4Dt}, \\
x^i &\rightarrow& x^i + \delta^{ij} \delta x_{|j} \label{4Dx},
\eeq
induces the following transformation on the perturbed energy momentum tensor
\label{[PCGI 21, p 8 a 15]} 
\beq \nonumber
\delta \rho_\MM &\rightarrow& \delta \rho_\MM - \dot{\rho}_\MM \delta t ,\\  \nonumber
\delta P_\MM &\rightarrow& \delta P_\MM - \dot{P}_\MM \delta t, \\ \nonumber
\delta q_\MM   &\rightarrow& \delta q_\MM + \left( P_\MM + \rho_\MM \right) \delta t, \\ \nonumber
\delta \pi_\MM &\rightarrow& \delta \pi_\MM.
\eeq
This means in particular that the perfect fluid condition (\ref{perfect}) and the adiabatic condition (\ref{adiab}) are gauge invariant statements. 
The perturbed 4D Einstein's equations
\beq \label{4Dein} 
\kqd \delta T^\MM_{\mu \nu} =\delta G^{(4)}_{\mu \nu}  , 
\eeq
 then lead to the four scalar equations
\beq
\label{EQM6_4D}
  \kqd \delta \rho_\MM &=&  -6 \frac{\dot{a}}{a} \left( \dot{\Psi} + \frac{\dot{a}}{a} \Phi \right) + \frac{2}{a^2} \Delta \Psi,\\
  \kqd \delta q_\MM &=&  -2 \dot{\Psi} - 2 \frac{\dot{a}}{a} \Phi  \label{EQM4_4D},\\
  \kqd \delta \pi_\MM &=& \frac{\Psi - \Phi}{a^2}  \label{EQM1_4D}, \\ 
 \kqd \delta P_\MM &=&   
 2\left(2  \frac{\ddot{a}}{a} + \frac{\dot{a}^2}{a^2} \right)\Phi + 2 \frac{\dot{a}}{a} \dot{\Phi} +  
2 \ddot{\Psi} + 6 \frac{\dot{a}}{a} \dot{\Psi} + \frac{2}{3 a^2} \Delta \left( \Phi - \Psi\right) .
  \label{EQM8_4D}
 \eeq
Those equations are respectively the $00$, $0i$, traceless part of $ij$, and trace of $ij$ components of  
the scalar part of Einstein's equations  (\ref{4Dein}).  While the perturbed energy-momentum conservation equation 
\beq \nonumber
D^\mu T^\MM_{\mu \nu} = 0,
\eeq 
leads to   
\beq
\label{EQM5_4D}
\delta \dot{\rho}_\MM + 3 \frac{\dot{a}}{a} ( \delta \rho_\MM + \delta P_\MM) + \frac{\Delta}{a^2} \delta q_\MM - 3 (P_\MM+\rho_\MM) \dot{\Psi}  = 0, \\
\delta \dot{q}_\MM+   3 \frac{\dot{a}}{a}  \delta q_\MM  + \delta P_\MM +\left( P_\MM + \rho_\MM \right) \Phi + \frac{2}{3}  \Delta \delta \pi_\MM  = 0 \label{EQM3_4D}.
\eeq
Those two last equations can also be derived from 
the previous four (\ref{EQM6_4D}-\ref{EQM8_4D}), 
thanks to Bianchi identities. 

\label{appD}
\subsection{Cosmological perturbations on the brane} \label{B2}
The perturbed 4D effective Einstein's equations (\ref{effein}) read
\label{(PCGI 1,p 20)} 
\beq \nonumber
\delta G^{(4)\mu}_\nu = -\delta {\cal E}^\mu_\nu + \kcq \delta \Pi^\mu_\nu, 
\eeq
with the scalar perturbations of $\Pi$  given by \cite{Bridgman:2001mc} 
\begin{eqnarray} \nonumber
\delta \Pi^0_0 &=& -\frac{1}{6} {\rho} \delta {\rho},\\ \nonumber
\delta \Pi^i_j &=& \delta^i_j \delta \Pi_T +  \left(\Delta^i_j - 
\frac{1}{3} \delta^i_j \Delta\right) \delta \Pi_{TF}, \\ \nonumber
 &  \mbox{ with }& \left\{ \begin{array}{lcl} \delta \Pi_T &=& \frac{1}{6} \left(({\rho}+ {P}) \delta {\rho}+ {\rho} \delta {P} \right) \\ \nonumber
\delta \Pi_{TF} &=& -\frac{1}{12} ({\rho} + 3 {P}) \delta{{\pi}}\\ \end{array} \right.
\\  \nonumber
\delta \Pi^0_i &=& \frac{1}{6} \rho \delta q_{|i}.
\end{eqnarray}
With the notations of the text, this leads to the following perturbed effective Einstein's equations on the brane
\beq
 \label{EQM6sAPP} 
 \frac{\kcq}{6} \rho \delta \rho &=& - \kqd \delta \rho_\EE - \left\{
6 \frac{\dot{a}}{a} \left( \dot{\Psi} + \frac{\dot{a}}{a} \Phi \right) - \frac{2}{a^2} \Delta \Psi\right\}_\bb,  \\
 \frac{\kcq}{6} \rho \delta q  &=& -  \kqd \delta q_\EE -  \left\{ 2 \dot{\Psi} + 2 \frac{\dot{a}}{a} \Phi  \right\}_\bb \label{EQM4sAPP},\\
 \frac{\kcq}{12 }  \left( 3P + \rho\right) \delta \pi &=&  \kqd \delta \pi_\EE + \left\{\frac{\Phi - \Psi }{a^2}\right\}_\bb \label{EQM1sAPP}, \\
\frac{\kcq}{6}  \left((\rho+P)  \delta \rho + \rho \delta P \right) &=& - \kqd \delta P_\EE   + 
 \left\{\left(4  \frac{\ddot{a}}{a} + 2 \frac{\dot{a}^2}{a^2} \right)\Phi + 2 \frac{\dot{a}}{a} \dot{\Phi} +  
2 \ddot{\Psi} \right. \nonumber \\ && \left. \quad \quad \quad \quad \quad \quad 6 \frac{\dot{a}}{a} \dot{\Psi} + \frac{2}{3 a^2} \Delta \left( \Phi - \Psi\right) \right\}_\bb  \label{EQM8sAPP}.
\eeq
Those equations are respectively the $00$, $0i$, traceless part of $ij$, and trace of $ij$ components of  
the scalar part of the effective Einstein's equations  (\ref{effein}).  
While the effective matter energy-momentum conservation equation 
(\ref{consS}) and Weyl matter  energy-momentum conservation equation (\ref{consweyl}) lead to 
\beq
 \label{EQM5sAPP} \left\{
\delta \dot{\rho} + 3 \frac{\dot{a}}{a} ( \delta \rho + \delta P) + \frac{\Delta}{a^2} \delta q - 3 (P+\rho) \dot{\Psi} \right\}_\bb &=& 0, \\
\left\{ \delta \dot{q}+   3 \frac{\dot{a}}{a}  \delta q  + \delta P +\left( P + \rho \right) \Phi + \frac{2}{3}  \Delta \delta \pi  \right\}_\bb &=& 0
\label{EQM3sAPP},\\
 \left\{
\dot{\delta \rho_\EE} + 3 \frac{\dot{a}}{a} \left(\delta \rho_\EE + \delta P_\EE \right)  + \frac{\Delta}{a^2} \delta q_\EE \right\}_\bb
 &=&  0 \label{EQM7sAPP},\\
 \left\{ 
 \dot{\delta q_\EE} + 3 \frac{\dot{a}}{a} \delta q_\EE
+  \delta P_\EE + \frac{2}{3} \Delta \delta \pi_\EE \right\}_\bb&=&  \left\{ 
\frac{\kcq}{6 \kqd}(P+\rho) \right.\nonumber \\
&&  \left. \left( 3 \frac{\dot{a}}{a} \delta q + \Delta \delta \pi  -\delta \rho  \right) \right\}_\bb.\label{EQM9sAPP}
\eeq
Again, those four equations are not independent from the four previous ones thanks to Bianchi identities.

\section{Expression of the coefficients $\CC$}\label{expressC}
Here we give the expressions for the coefficients appearing in equations (\ref{PHI1}-\ref{DP1}). 

\beq
\CC^\Phi_{\Delta (0,0)} &=& 1+ \fU (\UU + 1) \nonumber, \\
\CC^\Phi_{ (1,0)} &=& 3 \{ 1+ \fU (\UU + 1)\}  \nonumber,\\
\CC^\Phi_{ (2,0)} &=& -3\fU (1+ \Upsilon) \nonumber, \\
\CC^\Phi_{ (0,1)} &=& 3  \{  \fU (\UU+2)( \UU+1) -\UU \}\nonumber ,\\
\CC^\Psi_{\Delta (0,0)}&=& 1 + \fU \nonumber, \\
\CC^\Psi_{ (1,0)}&=& 3(1+ \fU)  \nonumber,\\
\CC^\Psi_{ (2,0)}&=& - 3 \fU \nonumber,\\
\CC^\Psi_{ (0,1)}&=& 3  \fU (2+\UU)\nonumber,\\
\CC^\rho_{\Delta (0,0)} &=& \frac{\fU^2}{H^3}  \left( 2 + \UU  \right) \left\{ 12 H^3 + 10 \dot{H} H + \ddot{H} + \UU  \left( 8 H^3  + 2 \dot{H} H \right) + H^3 \UU^2 \right\}\nonumber ,\\
\CC^\rho_{\Delta^2 (0,0)} &=& \frac{\fU}{3 H^2}  \left(  6H^2 + 2 \hd  +3 H^2 \UU \right)\nonumber,\\
\CC^\rho_{(1,0)} &=& - \frac{3 \fU^2}{H^6} \left( 2 + \UU \right) \left\{ 
6 H^4 \hd + 7 H^2 \hd^2 + \hd^3 - H^3 \hdd  \right.\nonumber\nonumber \\
&&+ \left. \UU \left( 2 H^6 + 7H^4 \hd + 2 H^2 \hd^2 \right) + \UU^2 \left( H^6 + H^4 \hd \right) \right\}\nonumber,\\
\CC^\rho_{(2,0)} &=& - \frac{\fU^2}{H^4} \left( 2 + \UU \right)  \left\{ 3\left(6 H^4 + 5 H^2 \hd - \hd^2 + H \hdd \right) + 9H^4 \UU \right\}\nonumber, \\
\CC^\rho_{\Delta (2,0)} &=& - \fU \left( 2 + \UU \right)\nonumber, \\
\CC^\rho_{(3,0)} &=& 3  \fU  \left( 2 + \UU \right)\nonumber, \\
 \CC^\rho_{(0,1)} &=& - \frac{3 \fU^2}{H^4} \left( 2 + \UU \right) \left\{  
4 H^2 \hd + 6 \hd^2 - 2 H \hdd + \UU \left( 2 H^2 \hd + \hd^2 - H \hdd \right)\right\} \nonumber, \\ \nonumber
\CC^\rho_{\Delta (0,1)} &=&    \fU \left( 2 + \UU \right)^2\nonumber,\\
\CC^\rho_{ (1,1)} &=&  -3\fU \left( 2 + \UU \right)^2 \nonumber,\\
\CC^q_{\Delta(0,0)} &=& \frac{\fU^2}{H^4} \{24 H^4 + 20 H^2 \hd + 2 H \hdd + \UU (24 H^4 + 10 H^2 \hd - \hd^2 + H \hdd) + 6 \UU^2  H^4 \},\nonumber  \\  
\CC^q_{(1,0)} &=& -\frac{3 \fU^2}{H^6} (2+\UU)\{ \UU (2+ \UU) H^6 + (6 + 7 \UU + \UU^2 ) H^4 \hd + (7+2 \UU) H^2 \hd^2 + \hd^3 - H^3 \hdd\} \nonumber, \\
\CC^q_{\Delta(1,0)} &=& -\frac{\fU}{H^2} \{3H^2 (2+ \UU) +2 \hd\}  \nonumber, \\ 
\CC^q_{(2,0)} &=&  -\frac{3 \fU^2}{H^4} (2+\UU) \{ 3 (2+\UU) H^4 + 5 H^2 \hd - \hd^2 + H \hdd \} \nonumber, \\
\CC^q_{(3,0)} &=&  3 \fU (2+\UU)  \nonumber, \\
\CC^q_{(0,1)} &=&  -\frac{3 \fU^2}{H^4} (2+\UU)\{ 2 (2+ \UU) H^2 \hd + (6 + \UU) \hd^2 - (2+ \UU) H \hdd \} \nonumber, \\
\CC^q_{(1,1)} &=& -3 \fU ( 2 + \UU)^2, \nonumber \\
\CC^P_{\Delta(0,0)} &=& \frac{\fU^3}{H^8}\left\{ (2+\UU)^3 (-3+2\UU)H^6 \hd + (2+\UU) \hd^4 + 2 (2+\UU)^2 H^5 \hdd \nonumber \right.\\  && \left.
+ (28+20 \UU + 3 \UU^2) H^3 \hd \hdd - (4+ \UU) H \hd^2 \hdd + (2 + \UU) H^4 ((-24+ 5 \UU^2) \hd^2 \nonumber \right.\\  && \left.
- (2+ \UU) \hddd) - H^2 ((46+13 \UU - 2 \UU^2) \hd^3 - 2 (2+\UU) \hdd^2 + (2+ \UU) \hd \hddd) \right\} \nonumber, \\
\CC^P_{(1,0)} &=& \frac{\fU^3}{H^{10}}\left\{ 3 (2 \UU(2+ \UU)^3 H^{10} + (2 + \UU)^2(18 +25 \UU + 5 \UU^2) H^8 \hd - 3(-6 +\UU^2) H^2 \hd^4 \nonumber \right.\\  && \left.-\UU \hd^5 + (2+\UU)^2 (4 + 5\UU + \UU^2) H^7 \hdd + (44+52 \UU + 21 \UU^2 + 3 \UU^3) H^5 \hd \hdd \nonumber \right.\\  && \left. + (6+ 10 \UU + 3 \UU^2) H^3 \hd^2 \hdd + (2+\UU) H \hd^3  \hdd - 
(2+ \UU) H^6 ((-56-56\UU - 10 \UU^2 + \UU^3) \hd^2 \nonumber \right.\\  && \left. + (2+ \UU) \hddd) + H^4((42+ 39 \UU + \UU^2- 3 \UU^3) \hd^3 + 2(2+ \UU) \hdd^2 - (2+ \UU) \hd \hddd))\right\}_\bb \nonumber ,\\
\CC^P_{\Delta (1,0)} &=& -\frac{2 \fU^2}{ H^4} 
( 2 + \UU )\left\{ 3 \left( 2 + \UU \right) H^4 + 5 H ^2 \hd - \hd^2 + H \hdd  \right\}_\bb \nonumber, \\
\CC^P_{(2,0)} &=& \frac{\fU^3}{ H^8} \left\{3((2+ \UU)^3(6+\UU) H^8 + 2 (2+ \UU)^2 (13+4 \UU) H^6 \hd + \UU \hd^4\nonumber \right.\\  && \left. - 3(8+6 \UU + \UU^2) H^3 \hd \hdd + (4+ \UU) H \hd^2 \hdd
+ (2+ \UU) H^4 ((42+19\UU+ \UU^2) \hd^2\nonumber \right.\\  && \left. + (2+\UU) \hddd) + H^2 ((54+ 30 \UU+ 4 \UU^2) \hd^3 - 2 (2+ \UU) \hdd^2 + (2+ \UU) \hd \hddd))\right\}_\bb ,
\nonumber \\
\CC^P_{\Delta (2,0)} &=& \frac{\fU}{H^2} \left\{ 3(2+ \UU) H^2 + 2 \hd \right\}_\bb \nonumber ,\\
\CC^P_{ (3,0)} &=&  \frac{3 \fU^2}{H^4} \left\{ (2+ \UU)^2 H^4 + 3(2+ \UU) H^2 \hd - 2 (3+ \UU) \hd^2 + 2 (2+ \UU) H \hdd \right\}_\bb \nonumber, \\
\CC^P_{ (4,0)} &=& -3 \fU (2+ \UU) \nonumber, \\
\CC^P_{(0,1)} &=&  \frac{3\fU^3}{H^8} \left\{(2+ \UU)^3 (6+ \UU) H^6 \hd + (24+ 6 \UU - \UU^2) \hd^4 + (2+ \UU)^2(10+ 3 \UU) H^3 \hd \hdd\nonumber \right.\\  && \left. - (12+ 8 \UU + \UU^2) H \hd^2 \hdd - (2+ \UU)^2 H^4 ((-20 -2\UU + \UU^2) \hd^2 + (2+\UU) \hddd )\nonumber \right.\\  && \left. - (2+ \UU) H^2 ((-6+ 9 \UU + 4 \UU^2) \hd^3 - 2(2+ \UU) \hdd^2 + (2+ \UU) \hd \hddd) \right\}_\bb\nonumber ,\\
\CC^P_{(1,1)} &=& \frac{6 \fU^2}{H^4} (2 + \UU) \left\{(2+ \UU)^2 H^4 + 3(2+\UU) H^2 \hd + (6+\UU) \hd^2 - (2+ \UU) H \hdd \right\} \nonumber, \\
\CC^P_{(2,1)} &=&  3 \fU (2+ \UU)^2, \nonumber 
\eeq

where we have defined $\fU$ by 
\beq
\fU &=& \frac{1}{\Upsilon + \frac{2 H^2 + \dot{H}}{H^2}}. \nonumber
\eeq

\section{Initial data and well-poseness} \label{appDD}
We discuss here some issues related with 
the specification of initial data and  
the well-poseness of the
differential problem defined in sections \ref{BOUND} and \ref{MASTER}.
  We will assume that
initial data in the bulk are provided as indicated 
in the latter section and that a boundary condition on the
brane is given by equation (\ref{boundRS}) or (\ref{boundDGP}) of the former. 
We first discuss the relation between the bulk initial data and 
 initial data on the brane for real matter and induced metric
perturbations. The peculiarity of the boundary condition
(\ref{boundRS}) (or (\ref{boundDGP})) is that this boundary condition
involves derivatives of the master variable $\Omega$ (and of its normal derivative
$\Omega^\prime$) along the brane. In order
to shed light on this type of boundary condition,  we discuss
in subsection \ref{D2} a similar case where the PDE in the bulk is
simply a wave equation.

\subsection{Initial data}
Let us first consider the case where the initial data in the bulk are
specified on a curve which is non characteristic in the vicinity of
the initial event A chosen on the brane. This is the case for the
initial curve (AD) or (AEF) on  figure \ref{Fig1} (we refer
  here to figure \ref{Fig1} which concerns the DGP model,
  however the discussion also applies for the RS model).
In this case we specify initial data in the vicinity of A 
by giving Cauchy type of data. E.g. on (AE) we specify $\Omega$ and its
derivative normal to (AE). From these bulk data, one can get in A  all the partial derivatives of $\Omega$
of order greater or equal to  one
(this is because the initial data curve is non characteristic in A). This implies that the values of $\Omega$,
$\Omega^\prime$, and all their derivatives with respect to cosmic time 
 are known in A from the initial data in the bulk. These value should
 be compatible with the boundary condition (\ref{boundRS}) (or
 (\ref{boundDGP})). Furthermore this
 also implies that the values of the real matter and induced
 metric perturbations $\delta \rho_\MM,\delta P_\MM, \delta q_\MM,
 \Phi_\bb, \Psi_\bb$ are known initially in A from equations
 (\ref{PHIO}-\ref{DPO}) or (\ref{PHI1}-\ref{DP1}). 

Let us now turn to discuss the case where the initial curve is
characteristic in the vicinity of A (The problem has then some
similarities with the so-called Goursat problem). This is for exemple the case if
we specify initial data along the curve (AC) of figure \ref{Fig1}.
Since this is a characteristic curve we choose only $\Omega$ along it,
and can not specify freely there the  normal derivative of $\Omega$.   
This implies that the derivatives of $\Omega$ with respect to $y$ and
$t$ of order greater or equal to one are not determined in A from bulk
initial data. This is also true for the initial values of the real
matter and induced metric perturbations. However, one can specify in
A, in addition to characteristic initial data along (AC) (which allow
to get the values of $\partial^n \Omega / \partial T^n$, with $n\geq 1$, in A), the
derivatives $\partial^n  \Omega /\partial X^n$ (with $n\geq 1$).
If we do so, $\Omega$, $\Omega^\prime$, as well as their derivatives 
with respect to cosmic time along the brane, will be known in A.
This is also true for  the initial values of
real matter and induced metric perturbations. It is
interesting to note that we have here the same freedom, by specifying
$X-$derivatives in A,  to specify
initial data for real matter and induced metric as in the same case  
(i.e. adiabatic perturbations of a perfect fluid) in 
standard 4D cosmological perturbation theory.
Namely the specification in A of $\Psi_\bb$ and $\dot{\Psi}_\bb$ is
enough to obtain the initial values of the other perturbations.

\subsection{Well-poseness for the wave equation in the bulk}
In the simple case where the master equation is the wave equation 
\beq \label{wave}
\partial_X \partial_T \Omega =0, 
\eeq
here written in its characterstics coordinates, it is simple to see
that boundary conditions of the form (\ref{boundRS}) or
(\ref{boundDGP}) leads to a well posed problem, once initial data are
chosen as explained above. 
We only discuss here the case
where the curve on which we give initial data in the bulk is initially
non characteristic. We will for simplicity assume that the initial
data curve in the bulk is the curve (AEF) of figure \ref{Fig1}.
Let us then consider a given point P (which is in the future of A)
along the
brane trajectory in the two dimensional space $(X,T)$. We call $X_P$
and $T_P$ its coordinates. From the equation (\ref{wave}) we know that
$\partial_T \Omega$ is constant along the characteristic $T =
T_p$ (a similar discussion would hold for differential operators for
which one can build Riemann invariant along $T=constant$ characteristics). This means that $\partial_T \Omega$ is known at P from the
knowledge of the similar quantity at the intersection point Q of (AEF) and the
$T=T_p$ characteristic (the latter is given by initial data). From this we know then $\Omega^\prime$ in P as
a function of $\dot{\Omega}$, the brane trajectory and the initial data
in Q. Inserting this expression of $\Omega^\prime$ in the boundary condition (\ref{boundRS}) or
(\ref{boundDGP}) leads to an ordinary differential equation along the
brane for $\Omega$. The initial time derivatives of $\Omega$ needed to
  solve this equation are known from bulk initial data, as is
explained above. Then, solving for $\Omega$  along the brane, we get a
boundary condition along the brane which takes a familiar form. This
enables to conclude that the problem is indeed well posed in the case
considered here.

\label{D2}


\begin{thebibliography}{99} 

\bibitem{Arkani-Hamed:1998rs}
N.~Arkani-Hamed, S.~Dimopoulos and G.~R.~Dvali,
Phys.\ Lett.\ B {\bf 429} (1998) 263
[arXiv:hep-ph/9803315].
I.~Antoniadis, N.~Arkani-Hamed, S.~Dimopoulos and G.~R.~Dvali,
Phys.\ Lett.\ B {\bf 436} (1998) 257
[arXiv:hep-ph/9804398].
N.~Arkani-Hamed, S.~Dimopoulos and G.~R.~Dvali,
Phys.\ Rev.\ D {\bf 59} (1999) 086004
[arXiv:hep-ph/9807344].

\bibitem{Randall:1999vf}
L.~Randall and R.~Sundrum,
Phys.\ Rev.\ Lett.\  {\bf 83} (1999) 4690
[arXiv:hep-th/9906064].

\bibitem{DGP} G.~Dvali, G.~Gabadadze and M.~Porrati,
Phys.\ Lett.\  {\bf B485} (2000) 208
[arXiv:hep-th/0005016].

\bibitem{DG}G.~Dvali and G.~Gabadadze,
Phys.\ Rev.\ D {\bf 63}, 065007 (2001); [arXiv:hep-th/0008054].


\bibitem{Dvali:2001gm}
G.~R.~Dvali, G.~Gabadadze, M.~Kolanovic and F.~Nitti,
Phys.\ Rev.\ D {\bf 64} (2001) 084004
[arXiv:hep-ph/0102216].


\bibitem{Dvali:2001gx}
G.~R.~Dvali, G.~Gabadadze, M.~Kolanovic and F.~Nitti,
Phys.\ Rev.\ D {\bf 65} (2002) 024031
[arXiv:hep-th/0106058].




\bibitem{Witten:2000zk}
E.~Witten,
arXiv:hep-ph/0002297.



\bibitem{tot}
G.~R.~Dvali and S.~H.~Tye,
Phys.\ Lett.\ B {\bf 450} (1999) 72
[arXiv:hep-ph/9812483].
J.~Khoury, B.~A.~Ovrut, P.~J.~Steinhardt and N.~Turok,
Phys.\ Rev.\ D {\bf 64} (2001) 123522
[arXiv:hep-th/0103239].
P.~J.~Steinhardt and N.~Turok,
arXiv:hep-th/0111030.
M.~Bucher,
arXiv:hep-th/0107148.
U.~Gen, A.~Ishibashi and T.~Tanaka,
arXiv:hep-th/0110286.


\bibitem{cc} A.G. Riess et al., {\it Astroph. J} 116, 1009 (1998);\\
S.~Perlmutter {\it et al.}  [Supernova Cosmology Project Collaboration],
Astrophys.\ J.\  {\bf 517} (1999) 565
[arXiv:astro-ph/9812133].







\bibitem{Deffayet:2001uy}
C.~Deffayet,
Phys.\ Lett.\ B {\bf 502}, 199 (2001)
[hep-th/0010186].

\bibitem{Fifth}
C.~Deffayet, G.~R.~Dvali and G.~Gabadadze,
Phys.\ Rev.\ D {\bf 65} (2002) 044023
[arXiv:astro-ph/0105068].




\bibitem{Deffayet:2002sp}
C.~Deffayet, S.~J.~Landau, J.~Raux, M.~Zaldarriaga and P.~Astier,
[arXiv:astro-ph/0201164].

\bibitem{Avelino}
P.~P.~Avelino and C.~J.~Martins,
Astrophys.\ J.\  {\bf 565} (2002) 661
[arXiv:astro-ph/0106274].


\bibitem{comment}
C.~Deffayet, G.~Dvali and G.~Gabadadze,
astro-ph/0106449.


\bibitem{Veltman} H. van Dam and M. Veltman, Nucl. Phys. 
{\bf B22}, 397 (1970)~.
V.~I.~Zakharov, JETP Lett. {\bf 12}, 312 (1970)~.
 Y.~Iwasaki,
Phys.\ Rev.\ D {\bf 2} (1970) 2255.


\bibitem{LS}
J.~P.~Uzan and F.~Bernardeau,
Phys.\ Rev.\ D {\bf 64} (2001) 083004
[arXiv:hep-ph/0012011].
P.~Binetruy and J.~Silk,
Phys.\ Rev.\ Lett.\  {\bf 87} (2001) 031102
[arXiv:astro-ph/0007452].






\bibitem{GS}
J.~Garriga and M.~Sasaki,
Phys.\ Rev.\ D {\bf 62}, 043523 (2000)
[hep-th/9912118].
C.~Gordon and R.~Maartens,
Phys.\ Rev.\ D {\bf 63}, 044022 (2001)
[hep-th/0009010].
S.~W.~Hawking, T.~Hertog and H.~S.~Reall,
Phys.\ Rev.\ D {\bf 62}, 043501 (2000)
[hep-th/0003052].
K.~Koyama and J.~Soda,
Phys.\ Rev.\ D {\bf 62}, 123502 (2000)
[hep-th/0005239].
H.~A.~Bridgman, K.~A.~Malik and D.~Wands,
Phys.\ Rev.\ D {\bf 63} (2001) 084012 [hep-th/0010133].
N.~Sago, Y.~Himemoto and M.~Sasaki,
Phys.\ Rev.\ D {\bf 65} (2002) 024014
[arXiv:gr-qc/0104033].

\bibitem{Deruelle}
N.~Deruelle, T.~Dolezel and J.~Katz,
Phys.\ Rev.\ D {\bf 63}, 083513 (2001)
[hep-th/0010215];
N.~Deruelle and J.~Katz,
Phys.\ Rev.\ D {\bf 64} (2001) 083515
[arXiv:gr-qc/0104007].
N.~Deruelle and T.~Dolezel,
Phys.\ Rev.\ D {\bf 64} (2001) 103506
[arXiv:gr-qc/0105118].


\bibitem{Langlois}
D.~Langlois,
Phys.\ Rev.\ D {\bf 62}, 126012 (2000)
[hep-th/0005025];

\bibitem{Bridgman:2001mc}
H.~A.~Bridgman, K.~A.~Malik and D.~Wands,
Phys.\ Rev.\ D {\bf 65} (2002) 043502
[arXiv:astro-ph/0107245].


\bibitem{Langlois:2001ph}
D.~Langlois,
Phys.\ Rev.\ Lett.\  {\bf 86} (2001) 2212
[arXiv:hep-th/0010063].

\bibitem{Roy1}
R.~Maartens,
Phys.\ Rev.\ D {\bf 62}, 084023 (2000)
[hep-th/0004166];


\bibitem{LMSW}
D.~Langlois, R.~Maartens, M.~Sasaki and D.~Wands,
Phys.\ Rev.\ D {\bf 63} (2001) 084009 [hep-th/0012044].

\bibitem{Mukohyama:2000ui}
S.~Mukohyama,
Phys.\ Rev.\ D {\bf 62}, 084015 (2000)
[arXiv:hep-th/0004067].

\bibitem{Mukoh3}
S.~Mukohyama,
Class.\ Quant.\ Grav.\  {\bf 17} (2000) 4777 [hep-th/0006146];

\bibitem{Leong:2001qm}
B.~Leong, P.~Dunsby, A.~Challinor and A.~Lasenby,
arXiv:gr-qc/0111033.

\bibitem{cvdb}
C.~van de Bruck, M.~Dorca, R.~H.~Brandenberger and A.~Lukas,
Phys.\ Rev.\ D {\bf 62} (2000) 123515 [hep-th/0005032];
 C.~van de Bruck, M.~Dorca, C.~J.~Martins and M.~Parry,
Phys.\ Lett.\ B {\bf 495} (2000) 183 [hep-th/0009056];
 C.~van de Bruck and M.~Dorca,
hep-th/0012073;
M.~Dorca and C.~van de Bruck,
Nucl.\ Phys.\ B {\bf 605} (2001) 215
[arXiv:hep-th/0012116].
T.~Boehm, R.~Durrer and C.~van de Bruck,
Phys.\ Rev.\ D {\bf 64} (2001) 063504
[arXiv:hep-th/0102144].



\bibitem{Mukoh2}
S.~Mukohyama,
Phys.\ Rev.\ D {\bf 64} (2001) 064006
[arXiv:hep-th/0104185].


\bibitem{review}
D.~Wands,
arXiv:hep-th/0203107.
R.~Maartens,
arXiv:gr-qc/0101059.





\bibitem{Binetruy:2000ut}
P.~Binetruy, C.~Deffayet and D.~Langlois,
Nucl.\ Phys.\ B {\bf 565} (2000) 269
[arXiv:hep-th/9905012].





\bibitem{Shiromizu:2000wj}
T.~Shiromizu, K.~i.~Maeda and M.~Sasaki,
Phys.\ Rev.\ D {\bf 62} (2000) 024012
[arXiv:gr-qc/9910076].



\bibitem{Porrati:2002cp}
M.~Porrati,
arXiv:hep-th/0203014.


\bibitem{Deffayet:2002uk}
C.~Deffayet, G.~R.~Dvali, G.~Gabadadze and A.~I.~Vainshtein,
Phys.\ Rev.\ D {\bf 65} (2002) 044026
[arXiv:hep-th/0106001].

\bibitem{Lue:2001gc}
A.~Lue,
arXiv:hep-th/0111168.

\bibitem{Gruzinov:2001hp}
A.~Gruzinov,
arXiv:astro-ph/0112246.

\bibitem{Cline:1999ts}
J.~M.~Cline, C.~Grojean and G.~Servant,
Phys.\ Rev.\ Lett.\  {\bf 83} (1999) 4245
[arXiv:hep-ph/9906523].
C.~Csaki, M.~Graesser, C.~Kolda and J.~Terning,
Phys.\ Lett.\ B {\bf 462} (1999) 34
[arXiv:hep-ph/9906513].


\bibitem{Binetruy:2000hy}
P.~Binetruy, C.~Deffayet, U.~Ellwanger and D.~Langlois,
Phys.\ Lett.\ B {\bf 477} (2000) 285
[arXiv:hep-th/9910219].

\bibitem{Mukhanov:1992me}
J.~M.~Bardeen,
Phys.\ Rev.\ D {\bf 22} (1980) 1882.
V.~F.~Mukhanov, H.~A.~Feldman and R.~H.~Brandenberger,
Phys.\ Rept.\  {\bf 215} (1992) 203.
H.~Kodama and M.~Sasaki,
Prog.\ Theor.\ Phys.\ Suppl.\  {\bf 78} (1984) 1.


\bibitem{Deruelle:2000ge}
N.~Deruelle and T.~Dolezel,
Phys.\ Rev.\ D {\bf 62} (2000) 103502
[arXiv:gr-qc/0004021].

\bibitem{Ishihara:2001qe}
H.~Ishihara,
arXiv:gr-qc/0107085.


\bibitem{Mukohyama:1999wi}
S.~Mukohyama, T.~Shiromizu and K.~i.~Maeda,
Phys.\ Rev.\ D {\bf 62} (2000) 024028
[Erratum-ibid.\ D {\bf 63} (2000) 029901]
[arXiv:hep-th/9912287].

\bibitem{Kraus:1999it}
P.~Kraus,
JHEP {\bf 9912} (1999) 011
[arXiv:hep-th/9910149].

\bibitem{Giannakis:2002jg}
I.~Giannakis and H.~c.~Ren,
Phys.\ Lett.\ B {\bf 528} (2002) 133
[arXiv:hep-th/0111127].
R.~Dick,
Class.\ Quant.\ Grav.\  {\bf 18} (2001) R1
[arXiv:hep-th/0105320].




\bibitem{Deffayet:2001aw}
C.~Deffayet, G.~R.~Dvali, G.~Gabadadze and A.~Lue,
Phys.\ Rev.\ D {\bf 64}, 104002 (2001)
[arXiv:hep-th/0104201].

\end{thebibliography}
\end{document}